\documentclass[aps,pra,twocolumn,superscriptaddress,floatfix,nofootinbib,showpacs,longbibliography]{revtex4-2}

\usepackage[utf8]{inputenc}
\usepackage[T1]{fontenc}     
\usepackage[british]{babel}  
\usepackage[table]{xcolor}
\usepackage[scaled=0.86]{berasans}  
\usepackage[colorlinks=true, citecolor=blue, urlcolor=blue]{hyperref}
\usepackage{comment}
\makeatletter
\newcommand{\setword}[2]{%
  \phantomsection
  #1\def\@currentlabel{\unexpanded{#1}}\label{#2}%
}
\makeatother
\usepackage{comment}
\usepackage{graphicx} 
\usepackage[babel]{microtype}  

\usepackage{amsfonts}
\usepackage{amsmath}
\usepackage{amssymb,amsthm}

\usepackage{xspace}  
\usepackage{pgf,tikz}
\usepackage{xcolor}
\usepackage{multirow}
\usepackage{array}
\usepackage{bigstrut}
\usepackage{braket}
\usepackage{color}
\usepackage{natbib}
\usepackage{multirow}
\usepackage{mathtools}
\usepackage{dsfont}
\usepackage{float}
\usepackage[caption = false]{subfig}
\usepackage{xcolor,colortbl}
\usepackage{color}
\newcommand{\Tr}{\operatorname{Tr}}

\newcommand{\be}{\begin{equation}}
\newcommand{\ee}{\end{equation}}
\newcommand{\ba}{\begin{eqnarray}}
\newcommand{\ea}{\end{eqnarray}}

\newcommand{\tr}{\operatorname{Tr}}

\newtheorem{theorem}{Theorem}
\newtheorem{corollary}{Corollary}

\newtheorem{lemma}{Lemma}

\def\>{\rangle}
\def\<{\langle}

\usepackage{centernot}
\usepackage{subfig}
\usepackage{filecontents}

\providecommand{\ket}[1]{| #1{\rangle}}
\providecommand{\bra}[1]{\langle #1|}

\usepackage{newtx}

\begin{document}

\title{Ergotropic Characterization of Continuous-Variable Entanglement}

\author{Beatriz Polo-Rodríguez}
\affiliation{ICFO-Institut de Ciencies Fotoniques, The Barcelona Institute of Science and Technology, Av. Carl Friedrich Gauss 3, 08860 Castelldefels (Barcelona), Spain.}

\author{Federico Centrone}
\affiliation{ICFO-Institut de Ciencies Fotoniques, The Barcelona Institute of Science and Technology, Av. Carl Friedrich Gauss 3, 08860 Castelldefels (Barcelona), Spain.}
\affiliation{Universidad de Buenos Aires, Instituto de Física de Buenos Aires (IFIBA), CONICET,
Ciudad Universitaria, 1428 Buenos Aires, Argentina.}

\author{Gerardo Adesso}
\affiliation{School of Mathematical Sciences and Centre for the Mathematical and Theoretical Physics of Quantum Non-Equilibrium Systems, University of Nottingham, University Park, Nottingham, NG7 2RD, United Kingdom.}

\author{Mir Alimuddin}
\email{aliphy80@gmail.com}
\affiliation{ICFO-Institut de Ciencies Fotoniques, The Barcelona Institute of Science and Technology, Av. Carl Friedrich Gauss 3, 08860 Castelldefels (Barcelona), Spain.}

\begin{abstract} Continuous-variable quantum thermodynamics in the Gaussian regime provides a promising framework for investigating the energetic role of quantum correlations, particularly in optical systems. In this work, we introduce an entropy-free criterion for entanglement detection in bipartite Gaussian states, rooted in a distinct thermodynamic quantity: ergotropy—the maximum extractable work via unitary operations. By defining the \textit{relative ergotropic gap}, which quantifies the disparity between global and local ergotropy, we derive two independent analytical bounds that distinguish entangled from separable states. These bounds coincide for a broad class of quantum states, making the criterion both necessary and sufficient in such cases. Unlike entropy-based measures, our ergotropic approach captures fundamentally different aspects of quantum correlations and entanglement, particularly in mixed continuous-variable systems. We also extend our analysis beyond the Gaussian regime to certain non-Gaussian states and observe that Gaussian ergotropy continues to reflect thermodynamic signatures in entangled states, albeit with some limitations. These findings establish a direct operational link between entanglement and energy storage, offering an experimentally accessible approach to entanglement detection in continuous-variable optical platforms.
\end{abstract}

\maketitle	
{\it Introduction.--}
Quantum thermodynamics examines how intrinsically quantum resources shape energy and information processing in microscopic regimes \cite{QTBook,Goold2016,Vinjanampathy2016,Horodecki2013,Zurek2003,Brandao2013,Skrzypczyk2014,Popescu2006,Gour'15,PhysRevLett.134.050401,Friis_2016,Huber_2015,Vitagliano_2018}. With experimental advances enabling precise control over single quantum systems,  addressing the energetics of nanoscale devices has become both timely and essential \cite{Alexia}. A key concept in this context is {\em ergotropy}— the maximum amount of energy that can be extracted from a quantum system through unitary evolution~\cite{Allahverdyan2004}. This makes ergotropy a fundamentally operational and physically measurable quantity, capturing distinct features of quantum states that evade entropy-based analysis~\cite{Francica'20,Brais'20,Basu'25,Dominik2023,Francica2017,Mukherjee2016,Alimuddin2019,Alimuddin2020,Francica2022,Joshi2024,Puliyil2023,Yang2023,Akira'21,Simon'25,Yang'24,Alimuddin'25}.

Continuous-variable (CV) systems, particularly quantum optical platforms, provide a compelling setting for exploring these thermodynamic aspects. CV states are central to diverse quantum technologies, including quantum communication, sensing, and computation \cite{braunstein2005quantum,serafini2023quantum,adesso2014continuous}. Among these, Gaussian states and operations are particularly important due to their experimental feasibility \cite{weedbrook2012gaussian}, providing an attractive testbed for probing the energetic role of quantum correlations. Entanglement in Gaussian systems is typically assessed using entropic measures~\cite{adesso2007entanglement} or criteria such as the positive partial transpose (PPT) condition~\cite{lami2018gaussian,Yichen}. However, these information-theoretic tools lack direct connections to energy observables and may overlook operational features of quantum correlations.

This motivates the development of energy-based entanglement witnesses that operate independently of abstract mathematical metrics. Recent work suggests that ergotropy offers a thermodynamically meaningful framework for revealing quantum correlations~\cite{Francica2017,Mukherjee2016,Alimuddin2019,Alimuddin2020,Joshi2024,Puliyil2023,Yang2023,Akira'21,Simon'25,Yang'24,Alimuddin'25,Francica2022,Perarnau-llobet}. However, these ideas have primarily been explored in discrete-variable (DV) systems. Extending these concepts to CV systems is both challenging and essential, given their role in practical platforms like quantum batteries and optical processors~\cite{centrone2023charging}.

In this Letter, we introduce the \textit{relative ergotropic gap} (REG), which quantifies the disparity between extractable work from a global bipartite Gaussian state and its locally accessible components under Gaussian unitaries. We derive analytical bounds that distinguish separable from entangled states and show that REG is a necessary and sufficient indicator of entanglement in a wide class of Gaussian systems. Crucially, our analysis reveals that for mixed CV states, REG is fundamentally independent of entropic measures, demonstrating an unconventional approach to entanglement characterization. We further extend our energy-based framework to non-Gaussian states~\cite{Mattia}. In particular, for photon-subtracted states --- where the standard ergotropy is intractable due to their infinite-dimensional support --- we use the Shchukin–Vogel criterion~\cite{SV} to show that REG serves as a reliable entanglement witness in this class. These results highlight the utility of ergotropy as a robust witness of CV quantum correlations and open new directions for energy-driven quantum technologies.

{\it Energy Extraction in CV Systems.--} In thermodynamics, the work extractable from a system is controlled by its internal energy and entropy. For adiabatic processes, work extraction depends solely on internal energy changes. In the quantum regime, the analogous process is unitary evolution, which preserves von Neumann entropy. Ergotropy is the maximum amount of work reversibly extractable  \cite {Pusz1978,Lenard1978,Allahverdyan2004,Skrzypczyk2015}. For a state $\rho$ with Hamiltonian $H$, it is defined as $\mathcal{E}(\rho) := E(\rho) - \min_U E(U \rho U^\dagger),$ where the minimization is over all unitaries $U$, and $E(X) := \tr(X H)$ denotes the system's energy in state $X$. This expression quantifies the energy difference between $\rho$ and its passive counterpart $\rho^p$, the state with the same spectrum but minimal energy, beyond which no work can be extracted through unitaries. Ergotropy thus quantifies how far a state is from being passive, in purely energetic terms.

  In systems with infinite-dimensional Hilbert spaces, the implementation of a generic unitary is practically challenging.  However, 
  Gaussian states and operations form the backbone of optical quantum technologies, and their simplified mathematical structure makes them a natural playground for quantum thermodynamics in CV systems \cite{serafini2023quantum,brown2016passivity,Rodriguez'25,Serafini2020}. Restricting to Gaussian unitaries leads indeed to the concept of \textit{Gaussian ergotropy} \cite{brown2016passivity,Rodriguez'25}, defined as
\begin{equation}
\mathcal{E}_G(\rho) := E(\rho) - \min_{U_G} E(U_G \rho U_G^\dagger)= E(\rho) - E( \rho^p_G)
\end{equation}
where $U_G$ denotes a Gaussian unitary and $\rho^p_G$ is the corresponding {\it Gaussian passive state} that minimizes the system's energy under such operations. For an arbitrary state $\rho$, $\mathcal{E}_G(\rho) \leq \mathcal{E}(\rho)$, reaching the equality if $\rho$ is itself a Gaussian state, because Gaussian unitaries are optimal; in such cases, we omit the subscript \textit{G} and simply refer to ergotropy.

In composite or multi-mode CV systems, Gaussian ergotropy can be defined globally (via joint Gaussian unitaries $U_G$) or locally (via product unitaries $\otimes_i U_i$, where $U_i$ acts on $i^{th}$ mode). Their difference defines the \textit{Gaussian ergotropic gap}:
\begin{equation}\label{g-EG}
\begin{aligned}
\Delta \mathcal{E}_G &:= \mathcal{E}^{\text{global}}_G(\rho) - \mathcal{E}^{\text{local}}_G(\rho) =  E_{lp} - E_{gp}
\end{aligned}
\end{equation}
where $E_{lp}$ and $E_{gp}$ denote the energies of local and global Gaussian passive states. Operationally, this quantity captures the energetic advantage of performing correlated (global) operations over local ones. While the ergotropic gap has been studied in DV settings as a thermodynamic signature of quantum correlations \cite{Mukherjee2016,Alimuddin2019,Alimuddin2020,Puliyil2023}, its extension to CV systems remains unexplored.

In phase space, an $N$-mode bosonic system is described by the quadrature operators $\hat{\vec{r}} = (\hat{x}_1, \hat{p}_1, \ldots, \hat{x}_N, \hat{p}_N)^T$, which satisfy the canonical commutation relations
$ [\hat{r}_j, \hat{r}_k] = i \Omega_{jk}$,
with $\Omega = \bigoplus^N i\sigma_Y$ the symplectic form \cite{serafini2023quantum} and $\sigma_Y$ the Pauli $Y$ matrix.
The energy of a Gaussian state depends on its covariance matrix $\sigma$, with entries $
\sigma_{jk} = \langle \{ \hat{r}_j, \hat{r}_k \} \rangle - 2 \langle \hat{r}_j \rangle \langle \hat{r}_k \rangle$
and the system's Hamiltonian, which we assume in the form $\hat{H} = \sum_k \omega_k (\hat{x}_k^2 + \hat{p}_k^2)$. The mean energy of a Gaussian state $\rho$ with covariance matrix $\sigma$ is then
\begin{equation} \label{eq:gaussEnergy}
    E(\rho) = \mbox{$\frac{1}{4}$} {\sum}_k \omega_k \left( \text{Tr}[\sigma_k] - 2 \right),
\end{equation}
where $\sigma_k$ is the $2 \times 2$ covariance submatrix corresponding to mode $k$. Note that first moments also contribute to the energy of a Gaussian state, but since they can be removed through local unitary operations, they do not affect the ergotropic gap and thus can be disregarded in our analysis. For a two-mode Gaussian state, one can always bring the covariance matrix to the \textit{standard form} $\sigma_{sf}$ via local Gaussian unitary operations:
\begin{equation} \label{eq:standardFormCM}
\begin{aligned}
    \sigma_{sf} &= \begin{pmatrix}
        A & C \\
        C & B \\
    \end{pmatrix},
    \ \text{with } A= a \mathbb{I}, B= b\mathbb{I}, C = \begin{pmatrix}
        c_1 & 0 \\
        0 & c_2 \\
    \end{pmatrix}, \\ &\ a \geq b \geq 1 \text{ and } c_1 \geq |c_2|.
\end{aligned}
\end{equation}
This form is particularly useful for computing both local and global passive energies. The local passive covariance matrix $\sigma_{lp}$ is obtained via local Gaussian operations, while the global passive covariance matrix $\sigma_{gp}$ is achieved via a symplectic transformation that diagonalizes $\sigma_{sf}$ into its Williamson form: $\sigma_{gp} = \text{diag}(\nu_+, \nu_+, \nu_-, \nu_-)$, where $\nu_{\pm} \geq 1$ are the symplectic eigenvalues of $\sigma_{sf}$, i.e. the moduli of the eigenvalues of $i\Omega \sigma_{sf}$. Assuming $\omega_A \leq \omega_B$, the ergotropic gap for a two-mode Gaussian state $\rho_G$ reads:
\begin{equation}\label{eq:gap}
    \Delta \mathcal{E} := E_{lp} - E_{gp} = \mbox{$\frac{1}{2}$} [ \omega_A (a - \nu_+) + \omega_B (b - \nu_-) ]
\end{equation}
where these local and global passive energies are computed directly from the covariance matrices.

{\it Ergotropic Gap versus Correlations of Gaussian States.--} We begin by examining the interplay between quantum correlations and work extraction in {\it pure} two-mode Gaussian states.

\begin{theorem} \label{thm1}
    For pure two-mode Gaussian states
$\rho$, the ergotropic gap $\Delta \mathcal{E} (\rho)$
vanishes if and only if $\rho$ is separable. Moreover, $\Delta \mathcal{E} (\rho)$ is a strictly increasing function of the quantum mutual information $I(\rho)$, and hence provides an equivalent quantification of entanglement for these states.
\end{theorem}

The proof is discussed in Appendix B \cite{supp}. We find that the ergotropic gap for pure two-mode Gaussian states is fully characterized by a single parameter—just like all standard entanglement measures—keeping them mutually dependent. This observation closely parallels the case of pure $2 \times d$ dimensional discrete-variable (DV) systems, where all entanglement measures are functionally related to the entanglement entropy \cite{Vidal2000,Nielsen2000}. The result in Theorem~\ref{thm1} readily extends to pure bipartite Gaussian states of an arbitrary number of modes, by virtue of their mode-wise decomposition into products of two-mode states via local unitaries \cite{Botero}. In contrast, ergotropy-based entanglement measures in higher-dimensional or multipartite DV systems exhibit entropic independence \cite{Alimuddin2020,Puliyil2023}, arising from the presence of multiple independent Schmidt coefficients. Interestingly, despite the infinite-dimensional nature of the Gaussian regime, this entropic independence remains absent—highlighting a fundamental structural distinction between Gaussian and general DV systems for pure states.

This leads to a compelling question: can the ergotropic gap also serve as a faithful entanglement measure for {\it mixed} states? Moreover, since it arises from closed-system dynamics, does it capture a distinct notion of `quantumness' beyond what is reflected by conventional information-theoretic quantifiers? Before addressing this, we examine whether $\Delta \mathcal{E}(\rho)$ satisfies basic criteria for a correlation measure \cite{Divincenzo2004}. In DV systems, the ergotropic gap has been shown to witness entanglement, but it can be nonzero even for product states \cite{Alimuddin2019}. Interestingly, our next result reveals a distinct behavior in the Gaussian setting:

\begin{lemma}\label{lemma}
For general two-mode Gaussian states, $\Delta \mathcal{E}(\rho) > 0$ holds if and only if the state exhibits correlations.
\end{lemma}

The proof is Appendix D.1 \cite{supp}. This establishes that $\Delta \mathcal{E}(\rho)$ acts as a faithful indicator of correlations, being strictly positive if and only if the state is non-product. We now turn to the case of correlated mixed states, exploring whether this quantity can consistently reflect the strength of quantum correlations.

As a preliminary case study, we consider the well-known two-mode squeezed thermal state (TMS), obtained by applying a two-mode squeezing operation with parameter $z$ to two bosonic modes of frequency $\omega$, each initially in thermal equilibrium at temperature $T$. From the TMS covariance matrix, reported in Appendix C \cite{supp}, we find that the ergotropic gap $\Delta\mathcal{E}_{\text{TMS}}=2k\omega\sinh^2 r $ grows monotonically with temperature and diverges as $T \to \infty$. This is striking because thermal noise is generally expected to degrade quantum correlations. Indeed, the quantum mutual information (QMI) decreases with increasing temperature and remains bounded in the limit of infinite thermal noise: $ \lim_{k\rightarrow \infty} I(\sigma_{\text{TMS}})= 2 \log{\left(\cosh{\left(2 r \right)} \right)}$.

This behaviour of $\Delta \mathcal{E}$ underscores a fundamental distinction between CV and DV systems. In DV systems, rising temperature leads the state towards a maximally mixed configuration --- a global passive state --- causing the ergotropic gap to vanish. In contrast, in CV systems, the unbounded energy spectrum allows the thermal population of arbitrarily high energy levels without necessarily rendering the state passive. Consequently, even minimal correlations in extremely energetic states can lead to large ergotropic gaps despite diminishing QMI. This shows that the ergotropic gap $\Delta \mathcal{E}$, in its `vanilla' form, is not a suitable candidate for quantifying (classical or quantum) correlations in mixed CV states.

To mitigate the divergence problem, we introduce the {\em relative ergotropic gap} (REG), defined as
\begin{equation}\label{eq: REG}
   \Delta \mathcal{E}_{\text{rel}} \equiv \frac{\Delta \mathcal{E}}{E_{gp}} = \frac{E_{lp} - E_{gp}}{E_{gp}}
\end{equation}
where $E_{gp}$ denotes the energy of the global passive state. First, Lemma \ref{lemma} guarantees that REG is a faithful signature of (total) correlations. Second, as discussed in Appendix D.2 \cite{supp}, REG can also be regarded as a faithful witness of more general quantum correlations, like {\it discord}. Next, we investigate whether this quantity can capture the finer structure of quantum correlations, namely {\it entanglement} in the CV regime.

To this end, we first derive an explicit expression for $\Delta \mathcal{E}_{\text{rel}}$ for a general two-mode Gaussian state.  We decompose Gaussian operations acting on an initial product state of two thermal modes with temperatures $T_A$ and $T_B$ in terms of typical parameters of quantum optical systems \cite{braunsteinirreducible}. The resulting expression for the REG, as a function of the decomposition parameters, $\Delta \mathcal{E}_\text{rel} = f(k,\gamma,\alpha,\theta,z_1,z_2)$,
is explicitly presented in Appendix D.1 \cite{supp}. Here, $k=\frac{k_A+k_B}{2}$ is the mean thermal fluctuation factor, $\gamma =\frac{k_A-k_B}{2}$ is the fluctuation difference, $\alpha=\frac{\omega_B}{\omega_A}$ is the frequency ratio between the two modes, $z_A$ and $z_B$ are the respective squeezing parameters, and $\theta$ represents the beam splitter angle applied to entangle the modes.

Having introduced the REG, we now reveal a key structural distinction with respect to QMI:
\begin{lemma}
The relative ergotropic gap $\Delta \mathcal{E}_{\textup{rel}}$ is functionally independent of the quantum mutual information.
\end{lemma}
This independence, demonstrated in Appendix F.1 \cite{supp}, highlights the intrinsic significance of REG and indicates that $\Delta \mathcal{E}_{\text{rel}}$ captures distinct features of quantum correlations not accounted by QMI alone. To further investigate this discrepancy, we compare entanglement witnesses based on QMI and REG, finding instances where certain entangled states are detected by the latter but not by the former, and vice versa (see Appendix F.2 \cite{supp}). Motivated by this, we explore whether $\Delta \mathcal{E}_{\text{rel}}$ can discriminate between qualitatively distinct types of correlations, such as those in separable versus entangled states.

\begin{theorem}\label{main}
     Let $\rho_G$ be a Gaussian state of two modes. \\
(a) If $\rho_G$ is separable, then $\Delta \mathcal{E}_{\text{rel}}(\rho_G)$ is upper-bounded by
\begin{equation}\label{main'a}
     \Delta \mathcal{E}_{\textup{rel}}(\rho_G) \leq  \mathcal{B}^{\textup{sep}}_{\mathrm{max}} (k, \gamma, \alpha)
\end{equation}
\textit{(b) If $\rho_G$ is entangled, then $\Delta \mathcal{E}_{\textup{rel}}(\rho_G)$ is lower-bounded by}
\begin{equation}\label{main'b}
     \Delta \mathcal{E}_{\textup{rel}}(\rho_G) >\mathcal{B}^{\text{ent}}_{\mathrm{min}}(k,\gamma, \alpha)
\end{equation}
\end{theorem}

\begin{figure}[h!]
    \centering
    \includegraphics[width=\linewidth]{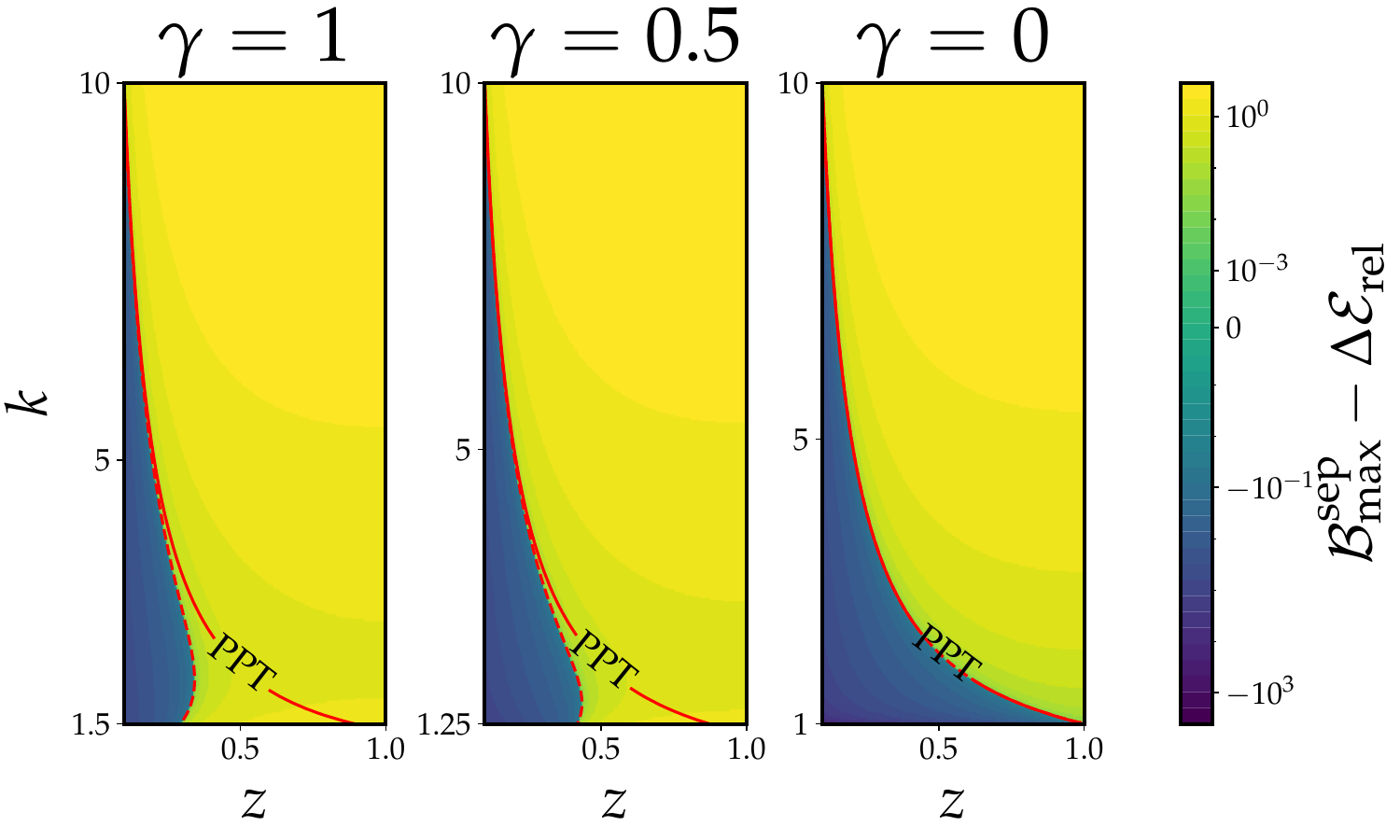}
    \caption{Difference between $\mathcal{B}^{\text{sep}}_{\mathrm{max}}$ and $\Delta \mathcal{E}_{\text{rel}}$ for the parametric family of Gaussian two-mode squeezed states, as a function of the squeezing parameter $z$ and the mean fluctuations factor $k$. Values of the fluctuation difference $\gamma$ have been set to \{1, 0.5, 0\} (left to right), and the frequency ratio is set at $\alpha=1$ for all cases. The dashed and solid red lines represent the boundaries between separable and entangled states according to our criterion [Eq.~(\ref{main'a})] and the PPT condition, respectively. As predicted by Corollary~\ref{corosame}, these boundaries coincide when $\gamma = 0$.}
    \label{fig:tms}
\end{figure}

The proof of the above theorem and explicit expressions for the bounds, together with a numerical illustration of the applicability of this result, are provided in Appendix E \cite{supp}. The two bounds are derived analytically employing the Positive Partial Transpose (PPT) criterion \cite{ppt_cv} as a necessary and sufficient condition for separability in Gaussian two-mode systems. Notably, they depend exclusively on the mean fluctuation parameters of the modes, their fluctuation difference, and the ratio of their respective frequencies. Recall that Lemma~\ref{lemma} establishes $\Delta \mathcal{E}_{\text{rel}}$ as strictly positive for all correlated states --- whether classical or quantum --- whereas Theorem~\ref{main} provides a sharper distinction via two analytic bounds: Eq.~(\ref{main'a}) as a witness for entanglement, and Eq.~(\ref{main'b}) as a witness for separability. These bounds satisfy
$\mathcal{B}^{\text{ent}}_{\mathrm{min}}(k, \gamma, \alpha) \leq \mathcal{B}^{\text{sep}}_{\mathrm{max}}(k, \gamma, \alpha)$, but they do not always coincide. As a result, $\Delta \mathcal{E}_{\text{rel}}$ is a useful {\it qualifier} but not a full entanglement quantifier  \cite{Horodecki2009}, since states in the `coexistence region' \cite{adesso2004,adesso2007entanglement} characterized by values of $\Delta \mathcal{E}_{\text{rel}}$ between the two bounds cannot be unequivocally classified.  However, the bounds do coincide in a physically relevant case.
\begin{corollary} \label{corosame}
Let $\rho_G$ be a Gaussian two-mode state such that the same fluctuation factor characterizes both modes. Then
\begin{equation}\label{eq: bound sep}
    \Delta \mathcal{E}_{\textup{rel}}(\rho_G) \leq \mathcal{B}^{\textup{sep}}_{\mathrm{max}}
\end{equation}
\textit{is a necessary and sufficient condition for separability.}
\end{corollary}
As a consequence of these results, the faithfulness of $\Delta \mathcal{E}_{\text{rel}}$ as an entanglement measure is determined by the distance between the bounds $\mathcal{B}^{\text{sep}}_{\mathrm{max}}$ and $\mathcal{B}^{\text{ent}}_{\mathrm{min}}$: the closer they are, the more accurately our criterion distinguishes between separable and entangled states. In the limiting case where the two bounds coincide, the condition based on the REG becomes equivalent to the PPT criterion. To illustrate this behavior, Figure~\ref{fig:tms} shows the degree to which $\Delta \mathcal{E}_{\text{rel}}$ violates the separability bound for the parametric family of two-mode squeezed Gaussian states TMS($k,z$), for different values of the fluctuation difference $\gamma$.  \\

{\it Non-Gaussian states.--} Gaussian states are both experimentally and theoretically important. However, their ease of characterization also comes with fundamental limitations regarding their usefulness in quantum information tasks \cite{Gaussbusters}. This motivates our exploration of the non-Gaussian regime \cite{Mattia} and the question of how far the relationship between the ergotropic gap and quantum correlations can be extended. For generic non-Gaussian states, evaluating the ergotropic gap becomes highly challenging and may even be intractable. By contrast, the Gaussian ergotropic gap $\Delta \mathcal{E}_G$ [Eq.~(\ref{g-EG})] --- together with the associated Gaussian relative ergotropic gap (G-REG) --- remains fully determined by the covariance matrix, thereby retaining computational accessibility. This observation suggests that while $\Delta \mathcal{E}$ provides the most general characterization, the Gaussian counterpart $\Delta \mathcal{E}_G$ may serve as a practical proxy in situations where the exact computation is prohibitive. In what follows, we investigate how $\Delta \mathcal{E}$ (where calculable), $\Delta \mathcal{E}_G$, and their relative forms can act as witnesses of entanglement in different families of non-Gaussian states.

By construction, the general results derived earlier (Theorems \ref{thm1} and \ref{main}) apply directly to non-Gaussian states obtained from Gaussian states through local unitary operations. However, the separable bound established in Eq.~(\ref{main'a}) cannot be straightforwardly generalized to arbitrary non-Gaussian states. To illustrate these subtleties, we now examine a specific class of states where $\Delta \mathcal{E}_G$ becomes trivial and fails as an entanglement witness, while $\Delta \mathcal{E}$ retains its detecting power.

\begin{lemma}
Any symmetric superposition of bipartite Fock states, i.e., any state of the form $\frac{1}{\sqrt{2}}(|n m\rangle + |m n\rangle)$ with $n,m \in \mathbb{N}$, is passive under Gaussian operations.
\end{lemma}

The proof, provided in Appendix G.1~\cite{supp}, shows that the covariance matrix of such states coincides with that of a Gaussian-passive state. Consequently, $\Delta \mathcal{E}_G=0$ even though the state is pure and entangled. Notably, however, the entanglement in this class can still be faithfully witnessed through $\Delta \mathcal{E}$, when ergotropy is defined with respect to arbitrary unitary operations.

Interestingly, there exist classes of non-Gaussian states where one can restore Gaussian non-passivity, leading to a nontrivial separable bound on $\Delta \mathcal{E}_G$. This is the case of the mixed bipartite Fock state $\rho = \lambda\ket{\phi^+}\bra{\phi^+}+(1-\lambda)\ket{\phi^-}\bra{\phi^-}$ with $n \in N$, where $\ket{\phi^{\pm}}= 1/\sqrt{2}(\ket{n,n}\pm \ket{n+1,n+1})$. For this example, $\rho$ is separable if and only if $\Delta \mathcal{E}_G = 0$ (See proof in Appendix G.2 \cite{supp}). This example highlights that within certain non-Gaussian families, both $\Delta \mathcal{E}$ and $\Delta \mathcal{E}_G$ can serve as necessary and sufficient entanglement witnesses.

Importantly, for the above class of Fock states,  $\Delta \mathcal{E}$ can be easily evaluated. In contrast, for genuinely infinite-dimensional non-Gaussian states, one must rely on $\Delta \mathcal{E}_G$ to circumvent both experimental and theoretical complications. This raises the question whether nontrivial separability bounds for $\Delta \mathcal{E}_G$ can be established in such cases. Below, we provide an affirmative answer by considering an experimentally relevant family: photon-subtracted TMS states~\cite{Mattia}.

\begin{figure}[h]
    \centering
    \includegraphics[width=\linewidth]{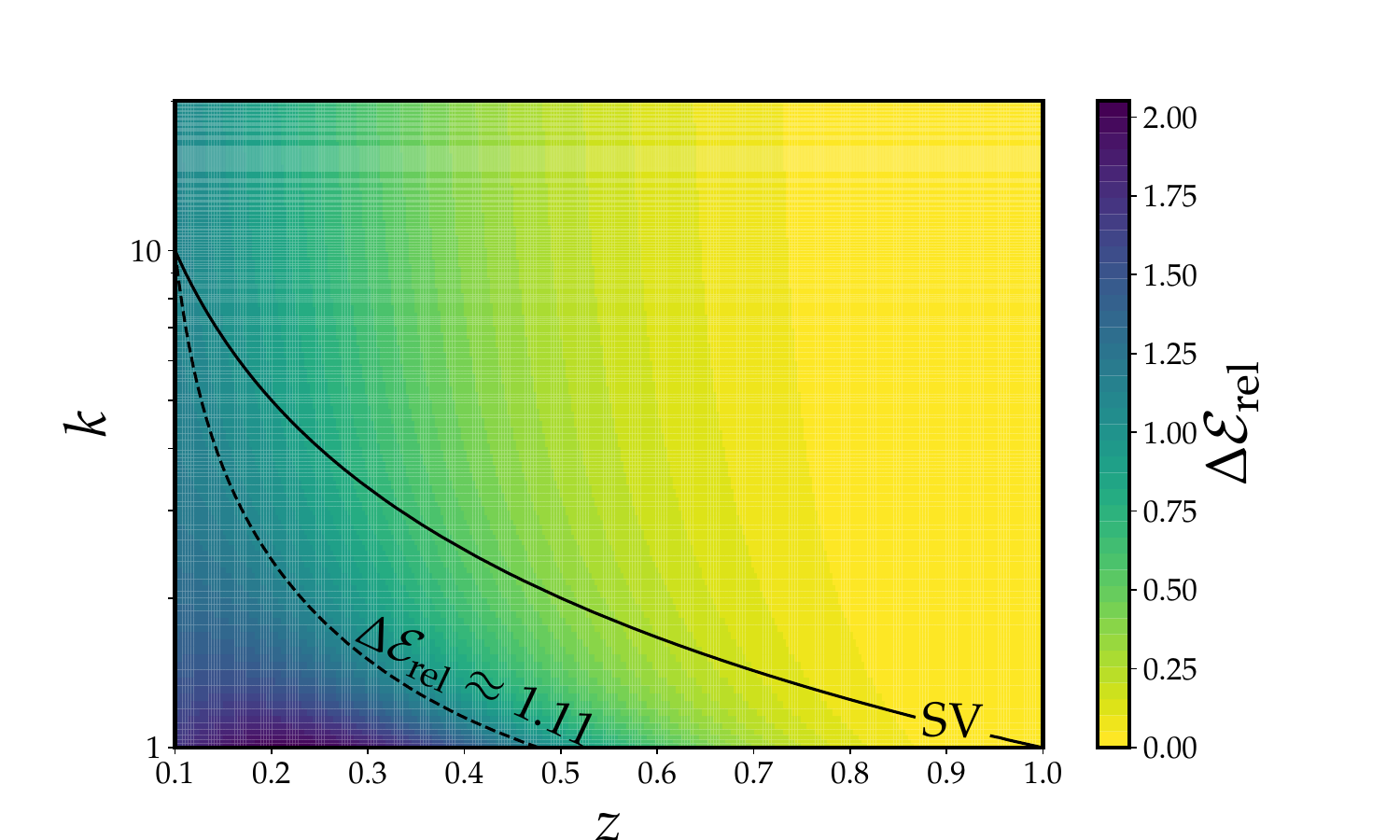}
    \caption{Gaussian relative ergotropic gap of photon-subtracted two-mode squeezed states as a function of the fluctuation parameter $k$ of the initial thermal state and the Gaussian squeezing parameter $z$. The region below the solid line is formed by entangled states as witnessed by the Shchukin-Vogel (SV) condition, while the region below the dashed line indicates entangled states witnessed by our G-REG-based criterion [Lemma~\ref{Lemmings}].}
    \label{fig: nongauss}
\end{figure}

Figure~\ref{fig: nongauss} shows $\Delta \mathcal{E}_{\text{rel}}$ as a function of the squeezing and fluctuation parameters $z$ and $k$ of the initial Gaussian TMS state prior to photon subtraction. Here, G-REG ($\Delta \mathcal{E}_G$) is computed using Eqs.~(\ref{eq:gaussEnergy}), (\ref{eq:gap}), and (\ref{eq: REG}), together with the covariance matrix derived in \cite{cm_photonsub}. As a separability benchmark we employ the Shchukin–Vogel (SV) criterion \cite{SV}, whose negativity suffices to certify entanglement. The maximum value of $\Delta \mathcal{E}_{\text{rel}}$ observed within the separable region (above the SV boundary) is then taken as a threshold, indicated by the dashed line in the figure.

\begin{lemma}For the non-Gaussian family of photon-subtracted two-mode squeezed states,
$\Delta \mathcal{E}_{\textup{rel}} \gtrapprox 1.11$ implies entanglement.\label{Lemmings}
\end{lemma}

A detailed proof is given in Appendix G.3 \cite{supp}. We emphasize that the separability bounds we establish for non-Gaussian states are inherently class dependent: a bound derived for one family cannot, in general, be used as an entanglement witness for another. It therefore remains an \textit{open problem} whether a universal nontrivial bound for all non-Gaussian separable states exists. If such a bound were found, G-REG could then serve as a fully general tool for entanglement witnessing in the non-Gaussian regime, independent of prior knowledge of the underlying state family.

{\it Discussion.--}
In this Letter, we characterize correlations in continuous-variable systems through thermodynamic processes. For bipartite Gaussian states, we demonstrate that correlation guarantees a strictly positive ergotropic gap, revealing an energy-based signature of correlations independent of entropy. Physically, this means that energy can be locked into correlations, making it inaccessible via local operations alone. For pure states, this yields an entanglement measure equivalent to the quantum mutual information. However, in mixed states, the ergotropic gap grows while correlations decrease due to the unbounded energy spectrum. To address this, we introduce the relative ergotropic gap (REG), a normalized quantity that is functionally independent of the quantum mutual information and more generally of entropic quantifiers. We derive analytical bounds on REG that distinguish entangled from separable states and serve as a necessary and sufficient separability criterion for a large class of states, paralleling the PPT test.
Notably, when energy measurements are available \cite{de2202experimental}, this could provide a more laboratory-friendly entanglement indicator than full state tomography. Moreover, we show that REG may find broader applicability outside the Gaussian regime --- an open direction for future study. Overall, our work establishes REG as an energy-based, experimentally accessible witness of continuous-variable quantum correlations, bridging thermodynamics and information-theoretic perspectives, and offering a novel tool for entanglement detection and characterization in near-term quantum optical technologies.

\begin{acknowledgements}
 {\bf Acknowledgements.--} This project has been funded by the Government of Spain (Severo Ochoa CEX2019-000910-S and FUNQIP), Fundació Cellex, Fundació Mir-Puig, Generalitat de Catalunya (CERCA program), and European Union (Quantera Veriqtas). F.C. acknowledges funding from the European Union (EQC, 101149233). G.A. acknowledges funding from the UK Research and Innovation (UKRI EPSRC Grant No. EP/X010929/1). M.A. acknowledges funding from the European Union (QURES, 101153001).

 The views and opinions expressed in this Letter are, however, the sole responsibility of the author(s).
\end{acknowledgements}

\textbf{Data availability.—}
The data supporting the findings of this study are available in Ref.~ \cite{Polo_Data}.

\onecolumngrid 

\appendix

\section{Separability of Gaussian two-mode states}\label{app:sep}
In the study of quantum information, the separability problem is a foundational question that addresses whether a given quantum state can be written as a convex mixture of product states. In this regard, the positive partial transpose (PPT) criterion \cite{ppt1,ppt2} is known to be a necessary condition for the separability of any bipartite quantum state, regardless of the Hilbert space dimension. In infinite-dimensional systems, the PPT condition can be evaluated on a bipartite state $\rho$ through its covariance matrix $\sigma$, yielding the following CV version of the criterion:
\begin{equation}
    \det(\sigma) - \det(\sigma_A) - \det(\sigma_B) + 2 \det(\sigma_{AB}) + 1 \geq 0
\end{equation}
where $\sigma_A$ and $\sigma_B$ are the local covariance matrices of the two subsystems $A$ and $B$ of the bipartition, and $\sigma_{AB}$ represents their correlations:
$$ \sigma = \begin{pmatrix} \sigma_A & \sigma_{AB} \\ \sigma_{AB} & \sigma_B \end{pmatrix} $$

In the context of Gaussian states, particularly in the two-mode case, it has been shown \cite{ppt_cv} that this condition is not only necessary but also sufficient for separability. Accordingly, if $\rho$ is a Gaussian state of two modes, and $\sigma_{sf}$ is the standard form of its covariance matrix: 

\[
\sigma_{sf}=  \begin{pmatrix}
    a & 0 & c_1 & 0 \\
    0 & a & 0 & c_2 \\
    c_1 & 0 & b & 0 \\
    0 & c_2 & 0 & b
    \end{pmatrix}, \quad  
 \sigma_A = a \mathbb{I}_2, \quad \sigma_B = b \mathbb{I}_2, \quad \sigma_{AB} = \begin{pmatrix} c_1 & 0 \\ 0 & c_2 \end{pmatrix},
\]

then the inequality 
\begin{equation} \label{separability_2modeGaussian_2}
    \det(\sigma_{sf}) - a^2 - b^2 + 2 c_1 c_2 + 1 \geq 0
\end{equation}
is satisfied if and only if $\rho$ is separable. Otherwise, it is entangled. \\

The applicability of this separability criterion extends beyond two-mode systems. In particular, it remains valid in the case of $1$ vs.\ $n$ mode bipartitions of Gaussian states \cite{ppt_cv}, as well as for \textit{bisymmetric} $(n+m)$ Gaussian states \cite{adesso2007entanglement} (i.e., states that are invariant
under mode permutations within any one of the two local subsystems). These generalizations provide a powerful framework for understanding entanglement structure in high-dimensional CV quantum systems while preserving computational tractability.

\section{Proof of Theorem 1}
The covariance matrix of a Gaussian two-mode pure state is characterized by a single parameter, and in standard form it reduces to Eq.4 taking $a=b$, $c_1= -c_2 = \sqrt{a^2 -1}$:
\begin{equation}
    \sigma_{\text{pure}}=\left[\begin{matrix}a & 0 & \sqrt{a^{2} - 1} & 0\\0 & a & 0 & - \sqrt{a^{2} - 1}\\\sqrt{a^{2} - 1} & 0 & a & 0\\0 & - \sqrt{a^{2} - 1} & 0 & a\end{matrix}\right],
\end{equation}
whereas through symplectic decomposition (i.e. diagonalization of $i\Omega\sigma_{\text{pure}}$) we obtain the eigenvalues $\nu_+ = \nu_{-} = 1$. Consequently, the ergotropic gap for these states reads
\begin{equation} \label{gap_purestates}
    \Delta \mathcal{E}_{\text{pure}} = \frac{\left(a - 1\right) \left(\omega_{A} + \omega_{B}\right)}{2}.
\end{equation}
Additionally, from Eq.~\ref{separability_2modeGaussian_2} one can straightforwardly derive the separability condition for these states:
\begin{equation}\label{separability_pure_states}
    \rho_{\text{pure}} \ \text{is separable} \iff 1 \geq a^2
\end{equation}
Since by definition $ 1 \leq a$, the only possibility for the RHS of Eq.~\ref{separability_pure_states} to be fulfilled is $a=1$, and thus the only 2-mode Gaussian pure separable (i.e product) states are those whose local passive state is the vacuum (and the standard form of their covariance matrix is the identity). From here and the strict possitivity of the frequencies $\omega_A, \omega_B$ it immediately follows that separability is a necessary and sufficient condition for $\Delta \mathcal{E}_{\text{pure}}=0$.  \\

Lastly, for the ergotropic gap to be a consistent entanglement quantifier, it must be proven to be monotonic under $LOCC$. It turns out this can be shown along with the functional dependence between $\Delta \mathcal{E}_{\text{pure}}$ and the mutual information $I(A:B)$ between the two modes:
\begin{equation}
    I(A:B) = (a+1) log_2 \bigg(\frac{a+1}{2} \bigg) - (a-1) log_2 \bigg(\frac{a-1}{2} \bigg) 
\end{equation} 
which is known to be an entanglement monotone for pure states \cite{Donald_2002}. Indeed, there exists a monotonically increasing function $f$ that uniquely maps $\Delta \mathcal{E}_{\text{pure}} \ \rightarrow \ I(A:B)$, namely

\begin{equation}
    f(x) =  (\gamma x+2) log_2 \bigg(\frac{\gamma x+2}{2} \bigg) - (\gamma x) log_2 \bigg(\frac{\gamma x}{2} \bigg)
\end{equation}
where we have taken $\gamma \equiv \frac{2}{\omega_{a} + \omega_{b}}$, and hence both metrics are \textit{LOCC} monotones and provide equivalent information on the correlations of any pure two-mode Gaussian state. \qed
\section{Parametrization of Gaussian two-mode mixed states}\label{app: bloch-messiah}
Any Gaussian state $\rho$ can be fully described (up to displacements, which do not affect the ergotropic gap and are thus ignored here) by its covariance matrix (CM).
In this appendix, we describe the so-called Bloch-Messiah decomposition \cite{braunsteinirreducible,serafini2023quantum} of the most general CM of a two-mode Gaussian state. This decomposition provides a physically insightful way of expressing Gaussian states in terms of fundamental operations: local phase shifts, local squeezings, and two-mode beam splitter transformations. In this decomposition, any given covariance matrix $\sigma$ can be written as:

\begin{equation} \label{eq: bloch-messiah}
\sigma = P(\varphi_A, \varphi_B) \ B(\theta)\ S(\sqrt{z_A}, \sqrt{z_B})  \ V(k_A, k_B) \ S^T(\sqrt{z_A}, \sqrt{z_B})\ B^T(\theta) \ P^T(\varphi_A, \varphi_B)
\end{equation}

In what follows, we will adopt the $xpxp$ convention for mode ordering for the definition of the matrices appearing in the previous expression; that is, we will take the quadrature vector as $\hat{\vec{r}} = (\hat{x}_A, \hat{p}_A, \hat{x}_B, \hat{p}_B)^T$ as opposed to the alternative ordering $\hat{\vec{r}} = (\hat{x}_A  \hat{x}_B, \hat{p}_A, \hat{p}_B)^T$. The matrix

\begin{equation}
V(k_A, k_B) = \begin{bmatrix}
k_A & 0 & 0 & 0 \\
0 & k_A & 0 & 0 \\
0 & 0 & k_B & 0 \\
0 & 0 & 0 & k_B
\end{bmatrix} \quad \quad \text{where} \ \
k_A \geq k_B \ \ \text{for} \ \ \omega_A \leq \omega_B
\end{equation}
is a diagonal matrix representing local thermal fluctuations. It encodes the symplectic eigenvalues of the state, which correspond to the effective thermal excitation levels of each mode.
Specifically, $k_A$ and $k_B$ denote the thermal fluctuation parameters of modes A and B, respectively. $V(k_A, k_B)$ is itself the covariance matrix of a product of two uncorrelated thermal states, with each symplectic eigenvalue $k_i$ ($i \in {A, B}$) related to the corresponding temperature $T_i$ and mode frequency $\omega_i$ by
\begin{equation}
    k_i = \coth \bigg( \frac{\omega_i}{2 K_B T_i}\bigg) \quad  i \in \{A,B\}
\end{equation}
where $K_{B}$ denotes the Boltzmann constant. \\

For convenience, we define the mean and difference of the thermal fluctuations as
\begin{equation}\label{change_variables}
k \equiv \frac{k_A + k_B}{2}, \qquad \gamma \equiv \frac{k_A - k_B}{2}
\end{equation}
Here, $k$ quantifies the average thermal noise (mean excitation) present in the system, while $\gamma$ captures the asymmetry between the modes.\\

Matrix $S(\sqrt{z_A}, \sqrt{z_B})$ represents local squeezing operations:

\begin{equation}
S(\sqrt{z_A}, \sqrt{z_B}) = \begin{bmatrix}
\sqrt{z_A} & 0 & 0 & 0 \\
0 & 1/\sqrt{z_A} & 0 & 0 \\
0 & 0 & \sqrt{z_B} & 0 \\
0 & 0 & 0 & 1/\sqrt{z_B}
\end{bmatrix}
\end{equation}

with $\sqrt{z_A}$ and $\sqrt{z_B}$ the squeezing parameters for modes $A$ and $B$, respectively. These parameters are defined in terms of the effective quadrature rescaling due to local unitary operations (such as those generated by a single-mode squeezer). \\

In this parametrization, beam splitting is responsible for entanglement generation. The matrix associated to this transformation is:
\begin{equation}
B(\theta) = \begin{bmatrix}
\cos\theta & 0 & \sin\theta & 0 \\
0 & \cos\theta & 0 & \sin\theta \\
-\sin\theta & 0 & \cos\theta & 0 \\
0 & -\sin\theta & 0 & \cos\theta
\end{bmatrix}
\end{equation}
which codifies the action of a beam splitter mixing the two modes, where $\theta \in [0, \pi/2]$ denotes the mixing angle, controlling the strength of mode entanglement. A value of $\theta = 0$ corresponds to no interaction, while $\theta = \pi/4$ yields maximal mixing.\\

Lastly, $P(\varphi_A, \varphi_B)$ represents local phase rotations acting on each mode:
\begin{equation}
P(\varphi_A, \varphi_B) = \begin{bmatrix}
\cos\varphi_A & \sin\varphi_A & 0 & 0 \\
-\sin\varphi_A & \cos\varphi_A & 0 & 0 \\
0 & 0 & \cos\varphi_B & \sin\varphi_B\\
0 & 0 & -\sin\varphi_B & \cos\varphi_B
\end{bmatrix}
\end{equation}

where $\varphi_A$ and $\varphi_B$ are the local phase shift parameters acting on modes $A$ and $B$, respectively. Since energy is invariant under local phase shifts, matrix $P(\varphi_A, \varphi_B)$ does not play a role in ergotropy-related considerations and its parameters do not influence our results. \\

The symplectic structure of the covaraince matrix does not depend on the physical frequencies of the modes. However, these frequencies enter the problem through the Hamiltonian and are necessary to define local energies and evaluate the ergotropic gap. Accordingly, we define:
\begin{center}

$\omega \equiv \omega_A$: the natural oscillating frequency of mode $A$ \\

$\alpha \equiv \frac{\omega_B}{\omega_A}$: the frequency ratio between the two modes, where $\alpha \geq 1$ since $\omega_A \leq \omega_B$.
\end{center}

While this appendix uses the Bloch-Messiah decomposition to express the Gaussian covariance matrix due to its physical transparency and modular structure, it is important to note that the results derived in the main text — particularly those concerning ergotropic gap bounds — are parametrization-independent. Any symplectic parametrization that captures the full Gaussian degrees of freedom would yield the same quantitative results. However, using the Bloch-Messiah form allows for a clear physical interpretation of each operation: local preparation (squeezing and thermal noise), mode interaction (beam splitting), and passive transformations (phase shifts).\\

For the example of two-mode squeezed (TMS) states, discussed in the main text to illustrate the genuinely distinct behaviour of Continuous variable (CV) and DV ones, the above parametrization yields the following form of the covariance matrix, which is itself in standard form:\\
$$\small\sigma_{\text{TMS}}=\left[\begin{matrix}k \cosh{\left(2 r \right)} \mathbb{I} & k \sinh{\left(2 r \right)} \sigma_z \\  k \sinh{\left(2 r \right)}\sigma_z & k \cosh{\left(2 r \right)} \mathbb{I} \end{matrix}\right], \,\,\text{where}  \begin{cases} k=\coth\left(\mbox{$\frac{\omega}{2T}$}\right) \\
{\color{blue} e^{-2r} = z}
\end{cases}$$

\section{Calculation of relative ergotropic gaps}
Following Eq.~6, the derivation of the relative ergotropic gap for a generic state $\rho$ reduces to finding its local and global passive energies. Moreover, since these energies depend only on first and second statistical moments (and first moments can be disregarded since they may be brought to zero through local Gaussian unitaries on the independent modes), the calculation of $\Delta \mathcal{E}_{\text{rel}}$ for any (not necessarily Gaussian) state depends solely on the entries of its covariance matrix. 

\subsection{Gaussian two-mode states \& Proof of Lemma 1}
Let $\sigma$ be the CM of a two-mode Gaussian state $\rho_G$, parametrized as explained in the previous Appendix. Then its corresponding global passive state is trivially that whose covariance matrix is 
\begin{equation}
\sigma_{\text{gp}} = \begin{bmatrix}
k_A & 0 & 0 & 0 \\
0 & k_A & 0 & 0 \\
0 & 0 & k_B & 0 \\
0 & 0 & 0 & k_B
\end{bmatrix}
\end{equation}
since this is the diagonal matrix of symplectic eigenvalues of $\sigma$. The global unitary that takes us from $\rho_G$ to its global passive is simply the inverse of the Gaussian transformations that define the parametrization of $\sigma$ (see Eq.~\ref{eq: bloch-messiah})

\begin{equation}
    \sigma_{\text{gp}} = G \ \sigma G^T \
\end{equation}
where 
\begin{equation}
    G = \bigg(P(\varphi_A, \varphi_B) \ B(\theta)\ S(\sqrt{z_A}, \sqrt{z_B}) \bigg)^{-1}
\end{equation}
and $\sigma$ is defined as in Eq.~\ref{eq: bloch-messiah}. The global passive energy can be straightforwardly computed from the symplectic eigenvalues and Eq.~3:
\begin{equation}
    E_{gp} = \frac{\omega}{2} [k_A + \alpha k_B - (1+\alpha)] = \frac{\omega}{2}[(k-1)(1+\alpha) +\gamma (1-\alpha)]
\end{equation}\\

For the local passive energy, we recall that any state whose covariance matrix is in standard form is locally passive. Therefore, our job reduces to finding the local Gaussian (symplectic) transformation that transforms $\sigma$ into $\sigma_{sf}$. The most general local Gaussian transformation can be decomposed in the product of single-mode squeezing and local rotations as: 
\begin{equation}
    G_{l o c} = P(\eta_A, \eta_B) \ S (r_A, r_B) \ P(\beta_A, \beta_B) 
\end{equation}
When applying these local transformation $ G_{l o c} \ \sigma \ G_{l o c}^T$, we obtain the product of two local phase shifts together, $P(\beta_A, \beta_B)$ and $P(\varphi_A,\varphi_B)$, which can be absorbed into a single matrix (to avoid extra degrees of freedom), or even chosen so that $\beta_A=-\varphi_A, \beta_B= -\varphi_B$, since they are passive optics transformations that will not affect the energy. Same happens with the final phase shift $P(\eta_A, \eta_B)$, that will leave unchanged the energy of the state it is applied to, so we can also assume $\eta_A=0, \eta_B=0$. Finally, the standard form of $\sigma$ we are looking for takes the form:
\begin{equation} \label{sigmasf}
    \sigma_{sf} = S (r_A, r_B) \ \sigma  \ S (r_A, r_B)^T 
\end{equation}
with $r_A$, $r_B$ taking values such that the resulting product is effectively in standard form.
After some straight-forward algebra, one finds such form is only attained when
\begin{equation} \label{r1_opt}
    r_A = \bigg( \frac{k_Az_Bcos^2\theta + k_B z_A sin^2 \theta}{(z_A z_B) (k_Az_Acos^2\theta + k_B z_B sin^2 \theta)}\bigg)^{1/4}
\end{equation}
and
\begin{equation}\label{r2_opt}
    r_B= \bigg( \frac{k_Bz_Acos^2\theta + k_A z_B sin^2 \theta}{(z_A z_B) (k_Bz_Bcos^2\theta + k_A z_A sin^2 \theta)}\bigg)^{1/4}
\end{equation}
Substituting expressions \ref{r1_opt},\ref{r2_opt} and \ref{change_variables} into Eq.~\ref{sigmasf}, we derive the explicit covariance matrix of the local passive state, whose energy, computed through Eq.3, yields:
\begin{equation}
\resizebox{\hsize}{!}{$ E_{lp} = \frac{\omega}{2} \Bigg[
\sqrt{(k+\gamma)^2 \cos^4\theta +(k-\gamma)^2 \sin^4\theta +(k^2-\gamma^2)\cos^2\theta \sin^2\theta \bigg(\frac{z_1^2 + z_2^2}{z_1 z_2}\bigg) }  + \alpha \sqrt{(k-\gamma)^2 \cos^4\theta +(k+\gamma)^2 \sin^4\theta +(k^2-\gamma^2)\cos^2\theta \sin^2\theta \bigg(\frac{z_1^2 + z_2^2}{z_1 z_2}\bigg)} - (1+\alpha) \Bigg]$}
    \end{equation}
and combining the local and global passive energies, the analytical expression for the relative ergotropic gap of a general two-mode Gaussian state yields
\begin{equation}\label{eg mixed states}
\begin{aligned}
\Delta \mathcal{E}_{\text{rel}} =& \resizebox{0.95\hsize}{!}{$ \frac{1}{(k-1)(1+\alpha) + \gamma (1-\alpha)} \cdot   \Bigg[-\Bigg( k(1+\alpha) + \gamma (1-\alpha)\Bigg) +  \sqrt{(k + \gamma)^2 \cos^4 \theta + (k - \gamma)^2 \sin^4 \theta 
+ (k^2 - \gamma^2) \cos^2 \theta \sin^2 \theta \left( \frac{z_B^2 + z_A^2}{z_A z_B} \right)}$} \\
&\resizebox{0.6\hsize}{!}{$ + \alpha \sqrt{(k - \gamma)^2 \cos^4 \theta + (k + \gamma)^2 \sin^4 \theta 
+ (k^2 - \gamma^2) \cos^2 \theta \sin^2 \theta \left( \frac{z_A^2 + z_B^2}{z_A z_B} \right)} \Bigg]$}\\
\end{aligned}
\end{equation}\\ 

\textbf{The \textit{Proof of Lemma 1}} follows from the previous expression, where by inspection we conclude that the numerator of $\Delta \mathcal{E}_{\text{rel}}$, namely $\Delta \mathcal{E}$, vanishes is and only if $\theta=0$. In this case, there exist no correlations between the two modes of $\rho_G$, which can be written as a tensor product of the two. For any $\theta \in (0, 2\pi)$, i.e., in the presence of (not necessarily quantum) correlations, the ergotropic gap is non-vanishing $\Delta \mathcal{E} > 0$. \qed

\subsection{REG and Quantum Discord in the Gaussian Regime}

\textit{Lemma 1} established that the ergotropic gap acts as a faithful quantifier of total correlations. However, since this quantity diverges for alomost correlation free states, it does not qualify as a well-defined correlation quantifier.  By introducing the regularized form---the \emph{relative ergotropic gap} (REG)---we resolve this issue and obtain a faithful and well defined total correlation quantifier. Our central motivation, however, goes beyond total correlations: we seek a thermodynamically inspired, entropy-free criterion that can not only capture classical correlations but also reveal genuinely quantum features of a state. In particular, we ask whether REG can distinguish classical from quantum correlations, thereby serving as a witness of quantumness. 

Before characterizing quantumness, it is useful to recall how \textit{quantum correlation} is defined via the structure of the state space. Restricting to bipartite systems, any state can be classified as follows:

\begin{itemize}
    \item \textbf{Classical--Classical (CC)}: Both subsystems are diagonal in some orthonormal bases. \\
    $\rho_{cc} = \sum_i p_i \ket{\psi_i}\bra{\psi_i} \otimes \ket{\phi_i}\bra{\phi_i}$ (includes product states).
    \item \textbf{Classical--Quantum (CQ)}: The first subsystem is diagonal in an orthonormal basis; the second is not.
    \item \textbf{Quantum--Classical (QC)}: The second subsystem is diagonal in an orthonormal basis; the first is not.
    \item \textbf{Quantum--Quantum (QQ)}: Separable states that are neither CC, CQ, nor QC.
    \item \textbf{Entangled (ENT)}: States that are not separable.
\end{itemize}

A state in the CC class is entirely classical, as both subsystems consist of distinguishable states. This observation underlies the concept of \textit{quantum discord} \cite{Zurek2000,Ollivier2001,Henderson2001}, which captures quantumness beyond entanglement by measuring the difference in mutual information across local measurements. Quantum discord is inherently asymmetric: for example, if Alice’s subsystem lacks a classical description (i.e., is not diagonal in any orthonormal basis), then the discord from Alice to Bob ($D_A$) is non-zero.

The discord for the five categories of states is summarized below:

\[
\begin{array}{c|c}
\hline
\text{Class} & \text{Discord} \\
\hline
CC & D_A = 0,\ D_B = 0 \\
CQ & D_A = 0,\ D_B \neq 0 \\
QC & D_A \neq 0,\ D_B = 0 \\
QQ,\ ENT & D_A \neq 0,\ D_B \neq 0 \\
\hline
\end{array}
\]

Research in DV systems has explored the relationship between the ergotropic gap and discord \cite{Mukherjee2016,Francica2017,Alimuddin'25}. Specifically, \cite{Mukherjee2016} shows that for a non-degenerate Hamiltonian, the ergotropic gap $\Delta \mathcal{E}(\rho)$ is strictly positive for all non-CC states:

\[
\neg (\text{CC}) \Rightarrow \Delta \mathcal{E}(\rho) > 0.
\]

This leads to the following correspondence:

\[
\begin{array}{c|c|c}
\hline
\text{Class} & \text{Discord} & \text{Ergotropic Gap} \\
\hline
CC & D_A = 0,\ D_B = 0 & \mathcal{E}(\rho) = 0\ \text{or} > 0 \\
CQ & D_A = 0,\ D_B \neq 0 & \mathcal{E}(\rho) > 0 \\
QC & D_A \neq 0,\ D_B = 0 & \mathcal{E}(\rho) > 0 \\
QQ,\ ENT & D_A \neq 0,\ D_B \neq 0 & \mathcal{E}(\rho) > 0 \\
\hline
\end{array}
\]

Note that even product states in the CC category may exhibit a non-zero ergotropic gap in DV systems. However, this does not hold when the Hamiltonian is degenerate \cite{Alimuddin2019}.

We now extend this analysis to continuous-variable (CV) systems with \textbf{REG} as quantifier, focusing on \textbf{two-mode Gaussian states}. According to Adesso and Datta \cite{Adesso2010}, such states fall into only two categories:

\[
\begin{array}{c|c}
\hline
\text{Class} & \text{Discord} \\
\hline
CC\ (\text{Only product states}) & D_A = 0,\ D_B = 0 \\
QQ,\ ENT & D_A \neq 0,\ D_B \neq 0 \\
\hline
\end{array}
\]

Combining this classification with our Lemma 1, we obtain the following relationship for Gaussian states:

\[
\begin{array}{c|c|c}
\hline
\text{Class} & \text{Discord} & \text{Relative Ergotropic Gap} \\
\hline
CC\ (\text{Only product states}) & D_A = 0,\ D_B = 0 & \Delta \mathcal{E}_{\text{rel}}(\rho) = 0 \\
QQ,\ ENT & D_A \neq 0,\ D_B \neq 0 & \Delta \mathcal{E}_{\text{rel}}(\rho) > 0 \\
\hline
\end{array}
\]

In the Gaussian regime, the absence of classical correlations implies that \textbf{all non-product states exhibit quantum correlations}. Thus, $\Delta \mathcal{E}_{\text{rel}}(\rho) = 0$ only when the state is a product. Unlike DV systems, Gaussian product states \textbf{cannot} yield a non-zero ergotropic gap—a limitation imposed by the Gaussianity constraint. Non-Gaussian states, however, may exhibit such phenomena \cite{brown2016passivity}.

\textbf{In summary}, there is a robust connection between quantum discord and the ergotropic gap in Gaussian systems. While mutual information, discord, and the relative ergotropic gap each serve as valid indicators of quantum correlations, the ergotropic gap uniquely provides an \textit{energy-based characterization} that complements traditional entropy-based measures.

\section{Proof of Theorem 2 and Corollary 1}
Written in terms of the variables appearing in the Bloch-Messiah decomposition (Eq.~\ref{eq: bloch-messiah} with the substitution given by Eq.~\ref{change_variables}), the separability condition in Eq. \ref{separability_2modeGaussian_2} for two-mode Gaussian states reads
\begin{equation}\label{new_sep_cond}
    1 + k^4 - 2k^2 \gamma^2 + \gamma^4 - 2k^2 - 2\gamma^2  \geq 
4\cos^2 \theta \sin^2 \theta \bigg[\left( k^2 - \gamma^2 \right) \left( \frac{z_1^2 + z_2^2}{z_1 z_2} \right) - 2 (k^2 + \gamma^2) \bigg]
\end{equation}

\textit{Part (a).} First, we note that by expansion of the binomial squares $(k + \gamma)^2$ and $(k - \gamma)^2$, the sum of square roots in the expression of the relative ergotropic gap (Eq.~\ref{eg mixed states}) yields the following structure
\begin{equation}
    \sqrt{a+\delta}+\alpha\sqrt{a-\delta} \quad \text{with} \quad \delta = 2k\gamma (\cos^4\theta - \sin^4\theta)  
\end{equation}
Since $|\delta| \leq 2k\gamma$, we may construct the following chain of inequalities:
\begin{equation}
    \sqrt{a+\delta}+\alpha\sqrt{a-\delta} \leq (1+\alpha) \sqrt{a + |\delta|} \leq (1+\alpha) \sqrt{a + 2k\gamma} 
\end{equation}

and the relative ergotropic gap is upper bounded by 
\begin{equation}
    \Delta \mathcal{E}_{\text{rel}} \leq  \frac{-\Bigg[k(1+\alpha) + \gamma (1-\alpha)\Bigg] + (1+\alpha) \sqrt{(k^2 + \gamma^2) (\cos^4 \theta +  \sin^4 \theta )
+ (k^2 - \gamma^2) \cos^2 \theta \sin^2 \theta \left( \frac{z_2^2 + z_1^2}{z_1 z_2} \right) +2k\gamma}}{(k-1)(1+\alpha) + \gamma (1-\alpha)}
\end{equation}
Since $\cos^4 \theta +  \sin^4 \theta =1 - 2 \cos^2 \theta \sin^2 \theta$:
\begin{equation}
   \Delta \mathcal{E}_{\text{rel}} \leq \frac{-\Bigg[k(1+\alpha) + \gamma (1-\alpha)\Bigg] + (1+\alpha) \sqrt{k^2 + \gamma^2 + 2k\gamma 
+  \cos^2 \theta \sin^2 \theta \bigg[(k^2 - \gamma^2)\left( \frac{z_2^2 + z_1^2}{z_1 z_2} \right) - 2(k^2+\gamma^2)\bigg]}}{(k-1)(1+\alpha) + \gamma (1-\alpha)}
\end{equation}
Finally, if we assume $\rho$ is separable, then we can embed the separability condition (Eq.~\ref{new_sep_cond}) into the previous inequality, and we are left with
\begin{equation}
   \Delta \mathcal{E}_{\text{rel}}^{\text{sep}} \leq \frac{-\Bigg[k(1+\alpha) + \gamma (1-\alpha)\Bigg] + \frac{1+\alpha}{2} \sqrt{ 
 1 + k^4 + \gamma^4 +2k^2 + 2\gamma^2  - 2k^2 \gamma^2 + 8k\gamma }}{(k-1)(1+\alpha) + \gamma (1-\alpha)} \equiv \mathcal{B}^{\text{sep}}_{\mathrm{max}} \quad \qquad \qed
\end{equation} \\
\textit{Part (b).} Analogously to the previous proof, we proceed by merging the two square roots in Eq.~\ref{eg mixed states}, through the inequality 
\begin{equation}
\sqrt{a+\delta}+\alpha\sqrt{a-\delta} \geq (1+\alpha) \sqrt{a - |\delta|} \geq (1+\alpha) \sqrt{a - 2k\gamma} \qquad \text{since} \quad  |\delta| = |2k\gamma (\cos^4\theta - \sin^4\theta)| \leq 2k\gamma
\end{equation}
with this and the trigonometry identity $\cos^4 \theta +  \sin^4 \theta =1 - 2 \cos^2 \theta \sin^2 \theta$, one arrives at:
\begin{equation}\label{step2_converseproof}
\Delta \mathcal{E}_{\text{rel}} \geq \frac{-\Bigg[k(1+\alpha) + \gamma (1-\alpha)\Bigg] + (1+\alpha) \sqrt{k^2 + \gamma^2 - 2k\gamma 
+  \cos^2 \theta \sin^2 \theta \bigg[(k^2 - \gamma^2)\left( \frac{z_2^2 + z_1^2}{z_1 z_2} \right) - 2(k^2+\gamma^2)\bigg]}}{(k-1)(1+\alpha) + \gamma (1-\alpha)}
\end{equation}
Since Eq.~\ref{new_sep_cond} is an \textit{if and only if} condition for separability, it can also be applied to characterize entangled states. If $\rho$ is entangled (not separable), then necessarily Eq.~\ref{new_sep_cond} does not hold, and thus
\begin{equation}
     1 + k^4 - 2k^2 \gamma^2 + \gamma^4 - 2k^2 - 2\gamma^2  < 
4\cos^2 \theta \sin^2 \theta \bigg[\left( k^2 - \gamma^2 \right) \left( \frac{z_1^2 + z_2^2}{z_1 z_2} \right) - 2 (k^2 + \gamma^2) \bigg]
\end{equation}
Embedding this new inverse inequality into Eq.~\ref{step2_converseproof}, we finally obtain the lower bound on $\Delta \mathcal{E}_{\text{rel}}$ for entangled states:
\begin{equation}
   \Delta \mathcal{E}_{\text{rel}}^{\text{ent}} >\frac{-\Bigg[k(1+\alpha) + \gamma (1-\alpha)\Bigg] + \frac{1+\alpha}{2} \sqrt{ 
 1 + k^4 + \gamma^4 +2k^2 + 2\gamma^2  - 2k^2 \gamma^2 - 8k\gamma }}{(k-1)(1+\alpha) + \gamma (1-\alpha)} \equiv\mathcal{B}^{\text{ent}}_{\mathrm{min}} \quad \qquad \qed
\end{equation} \\

\begin{figure}[h]
    \centering
    \includegraphics[width=0.8\linewidth]{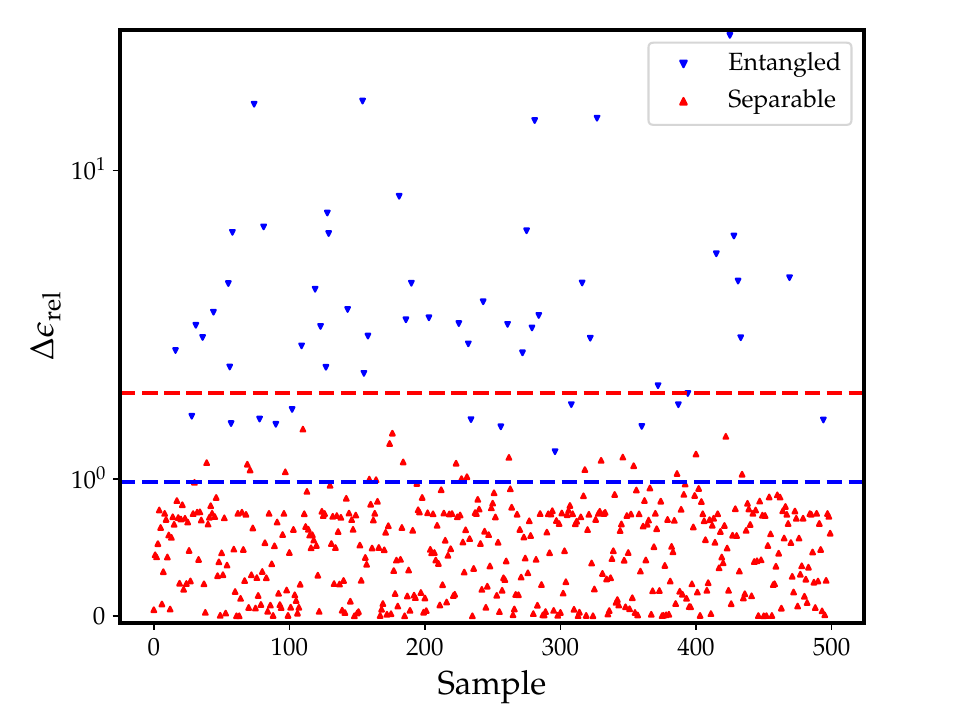}
    \caption{REG for $N=5 \times 10^2$ random two-mode Gaussian states with fixed parameters $k=2.5$, $\gamma=0.5$ and $\alpha=10$. The blue and red dashed lines represent the lower and upper bounds for entangled and separable states, respectively.}
    \label{bounds}
\end{figure}

Figure \ref{bounds} illustrates the meaning of these two bounds. Relative ergotropic gap is shown for $N=5 \times 10^2$ randomly generated two-mode Gaussian states. The blue dashed line represents the minimum $\Delta \mathcal{E}_{\text{rel}}$ for entangled states, while the upper bound for separable states is displayed in dashed red. Since the two bounds $\mathcal{B}^{\text{ent}}_{\mathrm{min}}$ and $\mathcal{B}^{\text{sep}}_{\mathrm{max}}$ are dependent on $k, \gamma$ and $\alpha$, the values of these parameters have been fixed so a single threshold holds for all of the simulated states. Neither entanglement nor separability of states with REG lying in between the two lines can be witnessed by means of our criterion. However, any value of REG above the red line or below the blue one can unambiguously certify that the state is entangled or separable, respectively.\\   

\textbf{Proof of Corollary 1:} \\
If $\rho_G$ has equal fluctuation factors $k_A,k_B$ for both modes, then $k = k_A = k_B$ and $\gamma = 0$.
In that case, $\mathcal{B}^{\text{sep}}_{\mathrm{max}} =\mathcal{B}^{\text{ent}}_{\mathrm{min}}$. From part \textit{(a)} of Theorem 2, we derive that Eq.9 is necessary for separability. Sufficiency follows from part \textit{(b)}.  \qed 
\section{Independency of relative ergotropic gap}\label{independency}

\subsection{Proof of Lemma 2}
We now demonstrate that, in contrast to the pure-state scenario, the \textit{relative ergotropic gap} remains a distinct and informative quantity for mixed states. In particular, it captures aspects of correlations not accounted for by the \textit{mutual information}. To assess whether these two quantities are functionally dependent, we analyze their gradients with respect to the set of variables appearing in the standard form of a generic covariance matrix \( \mathbf{x} = (a, b, \nu_+, \nu_-) \). Functional dependence between two scalar functions \( f(\mathbf{x}) \) and \( g(\mathbf{x}) \) can be tested by examining their \textit{Jacobian matrix}:

\begin{equation} \label{eq: jac}
J = \begin{bmatrix}
\frac{\partial f}{\partial a} & \frac{\partial f}{\partial b} & \frac{\partial f}{\partial \nu_+} & \frac{\partial f}{\partial \nu_-} \\ & \\
\frac{\partial g}{\partial a} & \frac{\partial g}{\partial b} & \frac{\partial g}{\partial \nu_+} & \frac{\partial g}{\partial \nu_-}
\end{bmatrix}
\end{equation}

If this matrix has \textit{full row rank} (i.e., rank 2) at a certain point, the functions \( f \) and \( g \) are functionally independent at that point in parameter space. We proceed by explicitly computing the gradients of both the \textit{relative ergotropic gap} \( \Delta \mathcal{E}_{\text{rel}} \) and the \textit{mutual information} \( I(A\!:\!B) \), and then evaluating the Jacobian matrix.\\

From Eqs.5 and 6, we derive the expression
\begin{equation}
    \Delta \mathcal{E}_{\text{rel}} = \frac{\omega_A (a - \nu_+) + \omega_B (b - \nu_-)}{\omega_A (\nu_+-1) + \omega_B(\nu_--1)}.
\end{equation}
We will assume $\omega_A = \omega_B$, since it does not affect the generality of the result. The corresponding gradient reads
\begin{equation}
    \nabla \bigg[\Delta \mathcal{E}_{\text{rel}} \bigg] =\bigg(\frac{\partial\Delta \mathcal{E}_{\text{rel}}}{\partial a}, \frac{\partial\Delta \mathcal{E}_{\text{rel}}}{\partial b}, \frac{\partial\Delta \mathcal{E}_{\text{rel}}}{\partial \nu_+}, \frac{\partial\Delta \mathcal{E}_{\text{rel}}}{\partial \nu_-}\bigg)= \bigg(\frac{1}{\nu_+ + \nu_- -2},\ \frac{1}{\nu_+ + \nu_- -2},\ \frac{ 2-b-a}{(\nu_+ + \nu_- - 2)^2}, \ \frac{ 2-b-a}{(\nu_+ + \nu_- - 2)^2} \bigg)  
\end{equation}
and for the mutual information, defined as
\begin{equation}
    I(A:B) = \frac{\sum_{x \in \{a,b\}} (x+1) log_2(\frac{x+1}{2}) -(x-1) log_2(\frac{x-1}{2}) - \sum_{y \in \{\nu_+, \nu_-\}}  (y+1) log_2(\frac{y+1}{2} ) - (y-1) log_2(\frac{y-1}{2})}{2}
\end{equation}
the gradient takes the form
\begin{equation}
     \nabla I(A:B) = \bigg(
    \frac{log_2(\frac{a+1}{2})-log_2(\frac{a-1}{2})}{2},
    \frac{log_2(\frac{b+1}{2})-log_2(\frac{b-1}{2})}{2},
    \frac{log_2(\frac{\nu_+-1}{2})-log_2(\frac{\nu_++1}{2})}{2},
    \frac{log_2(\frac{\nu_--1}{2})-log_2(\frac{\nu_-+1}{2})}{2}
    \bigg)
\end{equation}

By inspection of both gradients, one can immediately conclude they are independent vectors (it suffices to check that the first two entries of $\nabla \bigg(\Delta \mathcal{E}_{\text{rel}} \bigg)$ have the same value, whereas for $a\neq b$,
\begin{equation}
    \frac{log_2(\frac{a+1}{2})-log_2(\frac{a-1}{2})}{2} \neq
    \frac{log_2(\frac{b+1}{2})-log_2(\frac{b-1}{2})}{2}
\end{equation}
Consequently, the Jacobian matrix (see Eq.~\ref{eq: jac}) has rank 2, and the functions are proven to represent independent quantifiers of entanglement correlations for mixed states. It is important to note that the consideration of the particular case $\omega_A = \omega_B$ and $a\neq b$ in the above proof preserves the validity of the result, since functional independence of two functions at one single point in parameter space suffices to prove independence in general. \qquad \qed
\subsection{Comparison between entanglement witnesses}
As a consequence of the independence proven above, we now point out the existence of regions in the Gaussian parameter space where entanglement can be detected by our ergotropic criterion and not by witnesses based on mutual information or conditional entropy. We will examine the case of homogeneous frequencies ($\alpha=1$) and temperatures ($\gamma=0$) for simplicity. In such cases, the relative ergotropic gap of a two-mode Gaussian state reads
\begin{equation} \label{reg_homogeneous_temp_freq}
    \Delta \mathcal{E}_{\text{rel}} = \frac{k}{k-1} \bigg(\sqrt{\cos^4\theta + \sin^4 \theta + \cos^2 \theta \sin^2 \theta \bigg(\frac{z_A^2+z_B^2}{z_A z_B} \bigg) } -1 \bigg)
\end{equation}
and its corresponding bound for separable states yields
\begin{equation}\label{bound_homogeneous_temp_freq}
    \mathcal{B}^{\text{sep}}_{\mathrm{max}} = \frac{k^2-2k+1}{2 (k-1)}
\end{equation}
Defining $\tau \equiv \sqrt{\cos^4\theta + \sin^4 \theta + \cos^2 \theta \sin^2 \theta \bigg(\frac{z_A^2+z_B^2}{z_A z_B} \bigg) } $, the condition for entanglement
\begin{equation}
    \mathcal{B}^{\text{sep}}_{\mathrm{max}}(\rho) - \Delta \mathcal{E}_{\text{rel}}(\rho) < 0 \Longrightarrow  \rho \ \text{entangled}
\end{equation}
takes the form:
\begin{equation} \label{entanglementcondition_homogeneous_temp_freq}
    k^2 -2k\tau +1 <0 \ \iff \tau > \frac{k^2 +1 }{2k}\Longrightarrow  \rho \ \text{entangled}
\end{equation}
Meanwhile, the mutual information-based criterion states that any negative value of the conditional entropy serves as a sufficient condition for entanglement\cite{Horodecki1994}: 
\begin{equation}\label{QMIwitness}
    S (B | A)_\rho = S(\rho_{AB}) - S(\rho_A) <0 \Longrightarrow  \rho \ \text{entangled}.
\end{equation}
For the two-mode Gaussian states with equal frequencies and temperatures in both modes that we are considering, 
the covariance matrix of $\rho_{AB}$ has symplectic eigenvalues $\nu_{+} = \nu_{-}=k$, and that of $\rho_A$ has a single symplectic eigenvalue
$$a = k \sqrt{\cos^4\theta + \sin^4 \theta + \cos^2 \theta \sin^2 \theta \bigg(\frac{z_A^2+z_B^2}{z_A z_B} \bigg) } = k \tau.$$
The conditional entropy of these states can thus be computed through \cite{Wehrl1978}:
\begin{equation}
     S (B | A)_\rho = 2h(k) - h(k\tau) \quad \text{with} \quad h(x) = \frac{x+1}{2} \log_2 \bigg(\frac{x+1}{2} \bigg) - \frac{x-1}{2} \log_2 \bigg(\frac{x-1}{2} \bigg), 
\end{equation}
If we evaluate the function $2h(k) - h(k\tau)$ with $\tau > 1$ (since  $\tau > \frac{k^2 +1 }{2k}$ implies $\tau > 1 \ \text{for all} \  k \geq 1$), we obtain a monotonically increasing curve that intersects the $y$-axis and takes positive values in an entire open interval within $k \in [1,\infty)$. Consequently, for all the values of $k$ for which $ S (B | A)_\rho = 2h(k) - h(k\tau)$ is positive, we find entangled states spotted by our criterion and not by mutual information-based witnesses relying on the entropy of the state.

\subsection{A total summary and the role of QMI and REG }
In the following, we revisit the total correlation, as well as classical and different aspects of quantum correlation such as discord and entanglement, in terms of mutual information–based quantities\cite{Wilde2017}. Then, in the same light, we summarize how the ergotropic gap, or more specifically the relative ergotropic gap, can be used to characterize those correlations in a distinct way.

\medskip\noindent\emph{(1) Role of mutual information in classifying classical-quantum correlations.}
For a bipartite state $\rho_{AB}$,

$$
I(\rho)=S(\rho_A)+S(\rho_B)-S(\rho_{AB})=S(\rho_B)-S_\rho(B|A),
$$

where $S(\cdot)$ is the von Neumann entropy and $S_\rho(B|A)$ is the quantum conditional entropy. This quantity captures the \emph{total} correlations present (both classical and quantum). To quantify purely classical correlations, Henderson and Vedral \cite{Henderson2001} introduced the measurement-based quantity

$$
J_\rho^A = S(\rho_B) - \min_{M_A} S(\rho_B|M_A),
$$

which maximizes over local measurements on subsystem $A$. In general, one has

$$
I(X\!:\!Y)\leq J_\rho^A \leq I(\rho),
$$

where $I(X\!:\!Y)$ is the Shannon mutual information between optimal measurement outcomes on both parties. The difference $D_A=I(\rho)-J_\rho^A$ is the celebrated \emph{quantum discord} \cite{Henderson2001,Ollivier2001}, which captures more general nonclassical correlations.

While QMI is not a faithful entanglement measure (it assigns nonzero values to separable states), it can nevertheless \emph{witness} entanglement in some cases. Specifically, for separable states one always has $S_\rho(B|A)\geq 0$. Thus, any state with negative conditional entropy $S_\rho(B|A)<0$ necessarily satisfies $I(\rho)>S(\rho_B)$ and is entangled. This criterion is presented explicitly in Eq. \ref{QMIwitness}. Hence, QMI functions as a \emph{witness} of entanglement, though not as a {\it faithful} quantifier of quantum entanglement. We highlight this point because it is the conceptual template for how we use the regularized ergotropic gap (REG) in our work.

\medskip\noindent\emph{(2) The Gaussian ergotropic gap (EG): usefulness and limitations.}
The Gaussian ergotropic gap measures the extra extractable work available from a bipartite CV system under global Gaussian unitaries compared with local ones. Lemma 1 shows that EG is strictly positive whenever correlations (classical or quantum) are present, and vanishes only for product states.

However, to be a correlation quantifier, it needs to be well defined. EG fails in this respect and diverges even while the QMI decreases. This apparent paradox is important. The resolution is that the divergence in EG at high energies is not due to an increase in correlations but rather to the energy scaling of highly excited thermal states. In fact, both classical and quantum correlations, being upper bounded by QMI, necessarily decrease with temperature. Therefore, the growth of EG reflects a \emph{thermodynamic artifact} rather than genuine correlation content. This is why EG, in its raw form, is unsuitable as a correlation measure in mixed CV systems.

\medskip\noindent\emph{(3) Regularized ergotropic gap (REG): removing divergence and witnessing entanglement.}
To address this limitation, we define the \emph{regularized ergotropic gap} (REG), which removes the energy divergence and yields a well-behaved, correlation-sensitive quantity. REG vanishes on product states and grows smoothly with the strength of correlations, without unphysical divergence at high energies.

More importantly, we show that REG can also serve as an entanglement witness, in close analogy to QMI:
\begin{itemize}
    \item In Theorem 2, we prove that for all separable Gaussian states, the REG is upper bounded by a quantity explicitly given in Eq. (7) of the main text.
    \item Thus, if a state exhibits REG beyond this bound, the correlations must necessarily be of genuinely quantum origin, i.e., the state is entangled.
\end{itemize}

This is directly analogous to how negative conditional entropy allows QMI to witness entanglement. In fact, one may view REG as an \emph{energy-based} analogue of QMI in this sense. However, it is noteworthy that entanglement can act as a faithful quantifier of entanglement for a large class of states (see Corollary 1), where for separable states $\Delta \mathcal{E}_{\text{rel}}(\rho_G) \leq \mathcal{B}^{\text{sep}}_{\mathrm{max}}$ necessarily and sufficiently.

\medskip\noindent\emph{(4) Independence of REG and QMI.}
A crucial further point, captured in Lemma 2, is that REG is \emph{functionally independent} of QMI. This means that REG is not merely a disguised form of QMI but captures distinct structural features of correlations. Consequently, REG and QMI act as thermodynamically \emph{complementary} entanglement witnesses. There exist states detected by REG but not by QMI, and vice versa. For pure bipartite Gaussian states, where entanglement is characterized by a single parameter, both REG and QMI reduce to the same entanglement content (up to proportionality), fully consistent with Theorem 1.

\medskip\noindent\emph{(5) REG and discord.}
A natural concern is whether REG might simply be sensitive to classical correlations. To address this, in Appendix D.2 we analyze REG in the context of quantum discord. We show that REG captures not only entanglement but also the generic quantum correlations quantified by discord. In particular, even for separable states with nonzero discord, REG registers a nontrivial value, highlighting that it is genuinely sensitive to quantum features beyond classical correlations.

\section{Non-Gaussian states}
In this section we examine the possibility to extend the applicability of our Gaussian REG as a witness of entanglement for states beyond the Gaussian realm. 
The computation of the Gaussian REG for a non-Gaussian state is exactly analogous to that of a Gaussian state, and again requires knowledge of the state's covariance matrix (CM) exclusively, which contains the second statistical moments of the quadratures of the state. Such CM is not accessible in general, but it is known for some specific classes of non-Gaussian states. However, before moving directly to the Gaussian ergotropic scenario, we first examine non-Gaussian states for which the standard ergotropy remains tractable. In these regimes, it is important to understand how the standard ergotropy functions as an entanglement classifier and to compare its performance with the Gaussian ergotropy–based characterization.

\subsection{Symmetric superpositions of qudits}\label{app:qudits}

So far, we have described the framework in terms of CV quadratures. However, the harmonic oscillator Hamiltonian also admits finite-dimensional superpositions of Fock states. To make this connection, it is convenient to use the creation and annihilation operators, defined in terms of the quadratures as
\begin{equation} \label{eq: ladder}
    \begin{cases}
        \hat a=\frac{1}{\sqrt{2}}(\hat x+i\hat  p) \\[6pt]
        \hat a^{\dagger}= \frac{1}{\sqrt{2}}(\hat x-i\hat p)
    \end{cases}
\end{equation}
which satisfy the canonical commutation relation $[\hat a,\hat a^{\dagger}]=1$. The single-mode harmonic oscillator Hamiltonian then reads
\begin{equation}\label{harmonic oscillator}
   \hat  H=\omega\left(\hat a^{\dagger}\hat a+\tfrac{1}{2}\right).
\end{equation}

The eigenstates of $\hat H$, the Fock states $\ket{n}$, have a well-defined particle number, $\hat a^{\dagger}\hat a\ket{n}=n\ket{n}$. They are generated by successive applications of the creation operator, $\ket{n}=\tfrac{1}{\sqrt{n!}}(\hat a^{\dagger})^n\ket{0}$. The ladder operators act as $\hat a\ket{n}=\sqrt{n}\ket{n-1}$ and $\hat a\ket{0}=0$. The first moments of the quadrature vector $\hat{\vec{r}} = (\hat{x}, \hat{p})^T$ vanish for these states,
\begin{equation}
    \langle \hat{\vec{r}} \rangle_{\ket{n}}=\bra{n}\hat{\vec{r}}\ket{n}=(0,0)^T,
\end{equation}
as can be seen by expressing $\hat x,\hat p$ in terms of $\hat a,\hat a^{\dagger}$ and using the ladder operator algebra, which always connects orthogonal Fock states. \\

For two modes, one can define the eigenstates of $\hat H \otimes \hat H$ analogously as the tensor product of two single-mode Fock states, denoted by $|n_A, n_B \rangle$. In particular, as our first example of non-Gaussian states we will focus on their symmetric superpositions, i.e., 
\begin{equation}\label{eq: qudit symm superposition}
    |\psi \rangle = \frac{1}{\sqrt{2}} (|n,m\rangle + |m,n\rangle).
\end{equation}

For states of the form $|\psi\rangle$, the computation of the standard ergotropic gap is straightforward. 
Since $|\psi\rangle$ is a pure state, its global passive state coincides with the ground state of the Hamiltonian, namely 
$\psi_{0} = |0,0\rangle$, which has energy 
\[
E_{\mathrm{glob}}^{\mathrm{passive}} = 2\epsilon_0 =  \omega\!\left(\tfrac{1}{2} + \tfrac{1}{2}\right) = \omega
\]
for the two-mode system (see Eq.\ref{harmonic oscillator}).

To determine the local passive state, we first consider the reduced state of each subsystem obtained by tracing out the other mode. 
The reduced state is
\[
\rho_{\mathrm{A}} = \rho_{\mathrm{B}} = \tfrac{1}{2}\big(|n\rangle\langle n| + |m\rangle\langle m|\big),
\]
whose spectrum is 
\[
\{0,0,\dots,\tfrac{1}{2}^{(n)},0,\dots,\tfrac{1}{2}^{(m)}\}.
\]
By applying local unitary operations, we can rearrange the two largest eigenvalues so that they occupy the two lowest-energy levels, yielding the local passive state
\[
\rho_{\mathrm{loc}}^{\mathrm{passive}}  =\tfrac{1}{2}\big(|0\rangle\langle 0| + |1\rangle\langle 1|\big),
\]
whose energy is
\[
E_{\mathrm{loc}}^{\mathrm{passive}} 
   = \frac{1}{2}(\epsilon_0 + \epsilon_1) = \tfrac{1}{2}\omega + \omega + \tfrac{1}{2}\omega 
   = 2\omega .
\]

The standard ergotropic gap for this family of states is therefore
\[
\Delta \mathcal{E}_{\text{std}} = E_{\mathrm{loc}}^{\mathrm{passive}} - E_{\mathrm{glob}}^{\mathrm{passive}}
   = 2\omega - \omega
   = \omega ,
\]
which is independent of the specific values of $n$ and $m$. Note that if and only if $n=m$, the reduced states $\rho_{\mathrm{A}}, \rho_{\mathrm{B}}$ also become pure, with passive states equal to the ground state locally, each with energy $\epsilon_0$. In that case $E_{\mathrm{loc}}^{\mathrm{passive}} = E_{\mathrm{glob}}^{\mathrm{passive}}$ and the ergotropic gap vanishes. In short:
\begin{equation}
\Delta \mathcal{E}_{\text{std}} =
\begin{cases}
     0 & \text{if} \ n =  m \\
     \omega & \text{otherwise}
\end{cases}
\end{equation}
which turns $\Delta \mathcal{E}_{\text{std}}$ directly into a witness of entanglement (with the condition $\Delta \mathcal{E}_{\text{std}} > 0 \Longrightarrow \rho \ \text{entangled}$), since states of the form $|\psi\rangle$ in Eq.\ref{eq: qudit symm superposition} are trivially always entangled except when $n=m$. Through Lemma 3, we investigate whether the Gaussian REG is also able to detect the entanglement of this class of states, analogous to its standard counterpart. \\

\textbf{Proof of Lemma 3:}\\
To compute the Gaussian ergotropy of $|\psi\rangle$, it suffices to determine its covariance matrix, with entries $
\sigma_{kl} = \langle \{ \hat{r}_k, \hat{r}_l\} \rangle - 2 \langle \hat{r}_k \rangle \langle \hat{r}_l \rangle = \langle \hat{r}_k\hat{r}_l \rangle +  \langle \hat{r}_l\hat{r}_k \rangle $  (assuming zero first moments). From the definitions in Eq.~\ref{eq: ladder} and the commutation relation $[\hat a_i,\hat a^{\dagger}_j] = \delta_{ij}$, one can derive
\begin{equation}\label{eq: x-ladder identities}
\begin{cases}

    2 \hat{x}_i \hat{x_j}= \hat{a}_i \hat{a}_j +  \hat{a}_i \hat{a}^\dagger_j +  \hat{a}^\dagger_i \hat{a}_j +  \hat{a}^\dagger_i \hat{a}^\dagger_j \\
     2 \hat{p}_i \hat{p}_j= -\hat{a}_i \hat{a}_j +  \hat{a}_i \hat{a}^\dagger_j +  \hat{a}^\dagger_i \hat{a}_j -\hat{a}^\dagger_i \hat{a}^\dagger_j \\
     \hat{x}_i \hat{p_j} + \hat{p}_j \hat{x_i} = -i(\hat{a}_i \hat{a}_j -   \hat{a}^\dagger_i \hat{a}^\dagger_j + \hat{a}^\dagger_i \hat{a}_j - \hat{a}^\dagger_j \hat{a}_i)
    \end{cases}
\end{equation}
For this specific class of states (Eq.~\ref{eq: qudit symm superposition}), $\langle  \hat{a}^\dagger_i \hat{a}_i \rangle = \frac{n + m}{2}$ for all $i$, $\langle  \hat{a}^\dagger_i \hat{a}_j \rangle = \frac{m \delta_{n+1,m} + n \delta_{m+1,n}}{2} $ for all $i,j$ in the diagonal elements of the off-diagonal blocks; the rest of expectation values of pairs of ladder operators vanish. The only non-zero entries are thus the diagonal quadrature variances $\langle \hat{x_i}^2\rangle$ and $\langle \hat{p_i}^2\rangle$ , and the cross-terms of the form $\langle\hat{x}_i \hat{x_j}\rangle$ or $\langle \hat{p}_i \hat{p}_j\rangle$.\\

Denoting $a \equiv m + n +1$, and $c \equiv m \delta_{n+1,m} + n \delta_{m+1,n}$, we have the resulting covariance matrix for these states:
\begin{equation} \label{eq: CM symmetric sup of qudits}
    \sigma = \begin{pmatrix}
a & 0 & c & 0 \\
0 & a & 0 & c \\
c & 0 & a & 0 \\
0 & c & 0 & a
\end{pmatrix} = \begin{pmatrix}
a \mathbb{I}  & c \mathbb{I}  \\
c \mathbb{I} & a \mathbb{I} 
\end{pmatrix}
\end{equation}

The Gaussian ergotropy corresponds to

\begin{equation}
   \mathcal{E}_G= \frac{\omega}{4}(\Tr{\sigma}-2(\nu_--\nu_+)),
\end{equation}
where $\nu_{+},\nu_{-}$ are the symplectic eigenvalues of $\sigma$, and can be computed through
\begin{equation}
    \nu_{\pm}^2 = \frac{\Gamma \pm \sqrt{\Gamma^2 - 4\text{Det}\sigma}}{2} \quad, \quad \Gamma = \text{Det}\sigma_A + \text{Det}\sigma_B + 2\text{Det}\sigma_{AB}.
\end{equation}

With this, we can easily verify that $\mathcal{E}_G=0$ and as a consequence the matrix \ref{eq: CM symmetric sup of qudits} corresponds to the CM of a Gaussian passive state (a similar result can be found in Theorem 1 of \cite{brown2016passivity}). \qed \\

Consequently, no energy can be extracted from finite Fock-state symmetric superpositions of the form \ref{eq: qudit symm superposition} using Gaussian unitaries. This immediately implies that the Gaussian REG is not a valid entanglement witness in this regime, since it always vanishes. 

\subsection{Mixtures of Bell states}
Secondly, consider the finite-dimensional states belonging to the following parametrized family:
\begin{equation}\label{eq: bell mixture}
    \rho_\lambda =  \lambda |\phi^+\rangle \langle \phi^+| + (1-\lambda)|\phi^-\rangle \langle \phi^-|  \quad \quad \lambda \in [0,1]
\end{equation}
where 
\begin{equation}
    |\phi^+\rangle = \frac{1}{\sqrt{2}}(|n,n\rangle + |n+1,n+1\rangle), \quad \text{and} \quad |\phi^-\rangle = \frac{1}{\sqrt{2}}(|n,n\rangle - |n+1,n+1\rangle)
\end{equation}
It is known \cite{Horodecki'96} in literature that states in this class are entangled if and only if $\lambda \neq \frac{1}{2}$. We will now proceed to calculate their standard and Gaussian REG in order to see if they allow for a robust characterization of entanglement. \\

Writing \(\rho_\lambda\) in the basis of the two-dimensional subspace spanned by
\(|n,n\rangle\) and \(|n+1,n+1\rangle\) one finds
\[
\rho_\lambda=\tfrac{1}{2}\begin{pmatrix}1 & 2\lambda-1 \\[4pt] 2\lambda-1 & 1 \end{pmatrix},
\]
whose eigenvectors are \(|\phi^+\rangle\) and \(|\phi^-\rangle\) with eigenvalues
\(\lambda\) and \(1-\lambda\), respectively. The rest of its eigenvalues are 0. \\

The global passive state is obtained by a unitary that reorders the eigenvalues
of \(\rho_\lambda\) so that the largest eigenvalue occupies the lowest-energy eigenstate of the Hamiltonian (See Eq.\ref{harmonic oscillator}).
Therefore one assigns the larger of \(\{\lambda,1-\lambda\}\) to $|0,0\rangle\langle 0,0|$, which is the groundstate of $\hat{H}$,
and the smaller to $|0,1\rangle\langle 0,1|$. Hence the energy of the global passive state is
\begin{equation}
E_{\mathrm{glob}}^{\mathrm{passive}}
= \omega\bigg( \max(\lambda,1-\lambda)\ + 2\min(\lambda,1-\lambda)\bigg) = \omega \bigg(\min(\lambda,1-\lambda)+ 1 \bigg).
\end{equation}

Tracing out one mode to obtain the single-mode reduced state, we arrive at  
\[
\rho_{\text{A}} = \rho_{\text{B}}=\tfrac{1}{2}\big(|n\rangle\langle n| + |n+1\rangle\langle n+1|\big),
\]
for both \(|\phi^+\rangle\)
and \(|\phi^-\rangle\).

The spectrum of \(\rho_{A} = \rho_{B} \) is \(\{\,\tfrac{1}{2},\tfrac{1}{2}\,\}\) supported on levels \(n\) and \(n+1\).
By local unitaries we place these two populations on the two lowest single-mode energy levels, yielding the local passive state for each mode
\[
\rho_{\mathrm{loc}}^{\mathrm{passive}}=\tfrac{1}{2}\big(|0\rangle\langle 0| + |1\rangle\langle 1|\big).
\]
The energy of each mode in this state is
\[
E_{\text{mode}}^{\mathrm{passive}}=\tfrac{1}{2}\,\omega\big(\tfrac{1}{2}\big)+\tfrac{1}{2}\,\omega\big(\tfrac{3}{2}\big)=\omega,
\]
so the total local passive energy (two modes) is
\begin{equation}\label{Eloc}
E_{\mathrm{loc}}^{\mathrm{passive}} = 2\omega,
\end{equation}
which is independent of \(\lambda\) and \(n\). \\

The ergotropic gapy yields
\[
\Delta \mathcal{E}_{\text{std}} \;=\; E_{\mathrm{loc}}^{\mathrm{passive}} - E_{\mathrm{glob}}^{\mathrm{passive}}
 = 2\omega - \omega\bigg(1+ \min(\lambda,1-\lambda)\bigg)
         = \omega \bigg(1 -\min(\lambda,1-\lambda)\bigg) .
\]
By inspection of the above expression, we note that $\Delta \mathcal{E}_{\text{std}}$ has a minimum $\Delta \mathcal{E}_{\text{std, min}} = \frac{\omega}{2}$ at $\lambda=\frac{1}{2}$, which coincides with the only instance where $\rho_\lambda$ is separable. Thus, we find that $\Delta \mathcal{E}_{\text{std}} > \frac{\omega}{2}$ is a valid certification of entanglement. \\ 

With regard with the Gaussian REG, we begin by computing the covariance matrix of these states. In this case the only products of two ladder operators with non-vanishing expectation values are $\langle  \hat{a}^\dagger_i \hat{a}_i \rangle_{|\phi^{\pm}\rangle} =  \frac{2n + 1}{2}$, and $\langle  \hat{a}^\dagger_i \hat{a}^\dagger_j \rangle_{|\phi^{\pm}\rangle} = \langle  \hat{a}_i \hat{a}_j \rangle_{|\phi^{\pm}\rangle} = \pm \frac{n+1}{2} $. Applying the expressions in Eq.~\ref{eq: x-ladder identities}, the following covariance matrix is obtained as the linear combination of the covariance matrices of the two pure states in the mixture:
\begin{equation} \label{eq: bell mixture CM}
\sigma = \begin{pmatrix}
2(n+1) & 0 & |2\lambda -1|(n+1) & 0 \\
0 & 2(n+1) & 0 & -|2\lambda -1|(n+1) \\
|2\lambda -1|(n+1) & 0 & 2(n+1) & 0 \\
0 & -|2\lambda -1|(n+1) & 0 & 2(n+1)
\end{pmatrix} =
\begin{pmatrix}
a & 0 & c & 0 \\
0 & a & 0 & -c \\
c & 0 & a & 0 \\
0 & -c & 0 & a
\end{pmatrix} 
\end{equation}

Once the CM is known, the gap is computed exactly as if the states were Gaussian (see Eqs. 5 and 6 in the main text). It yields

\begin{equation}
    \Delta \mathcal{E}_G = \frac{E_{lp}- E_{gp}}{E_{gp}} = \frac{2a - \nu_{+} - \nu_{-}}{\nu_{+} + \nu_{-}-2},
\end{equation}
where the symplectic eigenvalues are
\begin{equation}
    \nu_{\pm}= (n+1) \sqrt{4-(2\lambda -1)^2},
\end{equation} 
and finally, the Gaussian relative ergotropic gap reads
\begin{equation}
    \Delta \mathcal{E}_G = \frac{ (2-\sqrt{4-(2\lambda -1)^2})}{\sqrt{4-(2\lambda -1)^2} - 1}.
\end{equation}
Interestingly and contrary to the previous example in DV, here we find that the Gaussian REG does work as an entanglement witness, since it is positive if and only if $\lambda \neq \frac{1}{2}$, i.e., if and only if $\rho$ is entangled.

\subsection{Photon-subtracted TMS states}
As our final example, we consider two-mode squeezed (TMS) states subjected to photon subtraction—a non-Gaussian operation—on the first mode. Such states are no longer supported on a finite-dimensional Hilbert space, which makes the direct calculation of their standard ergotropic gap computationally prohibitive. A common workaround is to truncate the Hilbert space and approximate the gap using the resulting discrete-variable (DV) states. Here, however, we address a different question, namely whether the Gaussian relative ergotropic gap (which, unlike the standard one, can be computed exactly) still serves as a meaningful indicator of the entanglement present in these continuous-variable (CV) non-Gaussian states.\\

\textbf{Proof of Lemma 4:}\\
The covariance matrix $\sigma$ of the photon-subtracted state $\rho$ is derived in \cite{cm_photonsub}:
\begin{equation}
\sigma = \sigma_0 + 2 \frac{(\sigma_0 - \mathbb{I}) P (\sigma_0 - \mathbb{I})}{\text{Tr}[(\sigma_0 - \mathbb{I}) P]}
\end{equation}

where $\sigma_0$ is the initial Gaussian TMS state's covariance matrix and $ P$ is the projection onto the subspace of the mode where the subtraction is performed. In our case (deterministic projection onto mode $A$):

\[
P = \begin{pmatrix}
1 & 0 & 0 & 0 \\
0 & 1 & 0 & 0 \\
0 & 0 & 0 & 0 \\
0 & 0 & 0 & 0
\end{pmatrix}
\]
The resulting $\sigma$ is in standard form (see Eq.~4), with:
\[
a= 
k\bigg(z+\frac{1}{z}\bigg)-1,  \quad  \quad b=\frac{k\bigg(z^2+\frac{1}{z^2}\bigg) - \bigg(z+\frac{1}{z}\bigg)}{z+\frac{1}{z}-\frac{2}{k}}, \quad  \quad c_1 = -c_2 = k\bigg(-z+\frac{1}{z}\bigg)
\]
and its symplectic eigenvalues can be computed through
\begin{equation} \label{eq: symplectic_eigvals}
    \nu_{\pm}^2 = \frac{\Gamma \pm \sqrt{\Gamma^2 - 4 \det \sigma}}{2}, \quad \Gamma = a^2 + b^2 + 2 c_1 c_2
\end{equation}

Finally, the expression for the Gaussian REG is directly derived from the above values and Eqs.~5 and 6. \\

As for the benchmark separability criterion, we employ one of the inequalities of the Shchukin-Vogel criterion \cite{SV}, whose negativity is sufficient for entanglement:
\begin{equation} \label{SV}
    \langle \hat a^\dagger \hat a\rangle_\rho \langle \hat b^\dagger \hat b\rangle_\rho -  
    \langle \hat a \hat b\rangle_\rho \langle \hat a^\dagger \hat b^\dagger \rangle_\rho  < 0 
    \ \Longrightarrow \ \rho \ \text{entangled},
\end{equation}
where $\hat a$ and $\hat b$ are the annihilation operators associated with modes $A$ and $B$, respectively. 

In order to establish our ergotropy-based condition, we identify the largest value of Gaussian REG within the region of states with $\langle \hat a^\dagger \hat a\rangle_\rho \langle \hat b^\dagger \hat b\rangle_\rho -  
    \langle \hat a \hat b\rangle_\rho \langle \hat a^\dagger \hat b^\dagger \rangle_\rho  \geq 0$, which is approximately $\Delta \mathcal{E}_{\text{rel, thres}} \approx 1.11 $. This value serves as a witness of entanglement for this class of states, since any larger value will lie within the area of entangled states detected by Eq.~\ref{SV}. In this sense, the condition G-REG $>\Delta \mathcal{E}_{\text{rel, thres}}$ is sufficient but not necessary for entanglement of photon-subtracted TMS states, and slightly weaker than the SV criterion (see Fig.~2 in main text for a comparison of the two bounds). \qed
%

\end{document}